\documentclass[twocolumn]{aastex631}

\usepackage{amsmath}
\usepackage{hyperref}

\usepackage{booktabs}
\usepackage{graphicx}

\tabletypesize{\scriptsize}

\received{17-Jan, 2024}
\revised{01-May, 2024}
\accepted{09-May, 2024}
\submitjournal{ApJ}

\shorttitle{A Spectroscopic Analysis of a Sample of K2 Planet-Host Stars}
\shortauthors{Loaiza-Tacuri et al.}

\graphicspath{{./}{figures/}}

\begin{document}

\title{Stellar Characterization and a Chromospheric Activity Analysis of a K2 Sample of Planet-Hosting Stars}

\correspondingauthor{Ver\'onica Loaiza Tacuri}
\email{vtacuri@on.br}

\author[0000-0003-0506-8269]{V. Loaiza-Tacuri}
\affiliation{Observat\'orio Nacional, Rua General Jos\'e Cristino, 77, 20921-400 S\~ao Crist\'ov\~ao, Rio de Janeiro, RJ, Brazil}

\author[0000-0001-6476-0576]{Katia Cunha}
\affiliation{Observat\'orio Nacional, Rua General Jos\'e Cristino, 77, 20921-400 S\~ao Crist\'ov\~ao, Rio de Janeiro, RJ, Brazil}
\affiliation{Steward Observatory, University of Arizona, 933 North Cherry Avenue, Tucson, AZ 85721, USA}

\author[0000-0002-0134-2024]{Verne V. Smith}
\affiliation{NSF's NOIRLab, 950 North Cherry Avenue, Tucson, AZ 85719, USA}

\author[0000-0001-8741-8642]{F. Quispe-Huaynasi}
\affiliation{Observat\'orio Nacional, Rua General Jos\'e Cristino, 77, 20921-400 S\~ao Crist\'ov\~ao, Rio de Janeiro, RJ, Brazil}

\author{Ellen Costa-Almeida}
\affiliation{Observat\'orio Nacional, Rua General Jos\'e Cristino, 77, 20921-400 S\~ao Crist\'ov\~ao, Rio de Janeiro, RJ, Brazil}

\author{Luan Ghezzi}
\affil{Universidade Federal do Rio de Janeiro, Observatório do Valongo, Ladeira do Pedro Antônio, 43, Rio de Janeiro, RJ 20080-090, Brazil}

\author{Jorge Melendez}
\affil{Instituto de Astronomia, Geofísica e Ciências Atmosféricas, Universidade de São Paulo, Rua do Matão 1226, São Paulo 05508-090, Brazil}

\begin{abstract}

Effective temperatures, surface gravities, and iron abundances were derived for 109 stars observed by the K2 mission using equivalent width measurements of Fe I and Fe II lines. Calculations were carried out in LTE using Kurucz model atmospheres. Stellar masses and radii were derived by combining the stellar parameters with Gaia DR3 parallaxes, V-magnitudes, and isochrones. The derived stellar and planetary radii have median internal precision of 1.8\%, and 2.3\%, respectively. The radius gap near $\rm R_{planet}\sim 1.9 R_\oplus$ was detected in this K2 sample. Chromospheric activity was measured from the Ca II H and K lines using the Values of $\log R^\prime_{\rm HK}$ were investigated as a function of stellar rotational period (P$_{rot}$) and we found that chromospheric activity decreases with increasing P$_{rot}$, although there is a large scatter in $\log R^\prime_{\rm HK}$ ($\sim$0.5) for a given P$_{rot}$. Activity levels in this sample reveal a paucity of F \& G dwarfs with intermediate activity levels (Vaughan-Preston gap). The effect that stellar activity might have on the derivation of stellar parameters was investigated by including magnetically-sensitive Fe I lines in the analysis and we find no significant differences between parameters with and without magnetically-sensitive lines, although the more active stars ($\log R^\prime _{\rm HK}>-5.0$) exhibit a larger scatter in the differences in $T_{\rm eff}$ and [Fe/H].

\end{abstract}


\section{Introduction} \label{sec:intro}
Currently there are $\sim$5500 confirmed exoplanets (NASA Exoplanet Archive) orbiting more than 4250 host stars, with the majority of detections resulting from planetary transits, the largest number of which were observed by the Kepler mission \citep{borucki2010,koch2010,borucki2016} and the extended Kepler mission, known as K2 \citep{howell2014}, as well as the CoRoT mission \citep{deleuil2000,deleuil2018}. The growing number of exoplanets has led to numerous studies whose main goals have been the characterization of planet-hosting stars and the planets orbiting them.  
Results from these studies have revealed fascinating patterns, such as the correlation between stellar metallicity and the frequency of giant planets \citep[][and subsequent studies]{gonzalez1997,fischer2005,Ghezzi2018}, the absence of planets having sizes between the super-Earth and sub-Neptune regimes \citep{fulton2017,fulton2018}, known as the radius valley, and a slope in the value of the radius valley as a function of planetary orbital period \citep[][among others]{vaneylen2018,Martinez2019}; such conclusions provide constraints on planetary formation theories \citep[e.g.][]{ida2004a,ida2004b,ida2005,nayakshin2010,mordasini2012,mordasini2015,owen2018,venturini2020}. 
The discovery of a planet-radius valley and its change as a function of orbital period was made possible thanks to high-precision spectroscopic methods used to determine the parameters and physical properties (effective temperature, T$_{\rm eff}$, surface gravity, as $\log g$, and metallicity, [Fe/H]) of the planet-hosting stars.  The derived planetary radii depend most strongly on the radii of their host-stars coupled with the transit depths, with the stellar radii derived from the stellar parameters, thus precise stellar characterization is critical to the derivation of precise planetary radii.

An additional stellar property that can impact derived stellar and planetary radii 
is magnetic activity on the host star \citep{yana2019MNRAS.490L..86Y,spina2020ApJ...895...52S}, as it can influence the characterization of the host star and, thus, the planets orbiting it. This influence arises from the connections between planetary properties, such as radius and mass, and those of their parent stars. The intensity of stellar magnetic activity can manifest itself in features such as spots and flares, or chromospheric UV, X-ray, and radio emission \citep{han2023ApJS..264...12H}. Emission lines such as the cores of Ca II H and K, Mg II, H-$\alpha$, the IR triplet of Ca II, and others are commonly used as indicators of chromospheric activity (or chromospheric emission). Among these indicators, the Ca II H ($\lambda$3968\AA) and K ($\lambda$3934\AA) lines, which originate in the photosphere, are commonly used as they are sensitive to chromospheric activity, which itself is associated with magnetic fields \citep{leighton1959ApJ...130..366L,skumanich1975ApJ...200..747S}.  When a star has an active chromosphere, the cores of these very strong Ca II absorption lines undergo reversals, due to the chromospheric temperature increase, and become core emission features which appear in sharp contrast to the deep photospheric absorption, making them excellent diagnostic tools for assessing chromospheric activity.

In this study we derive stellar parameters for a sample of planet-hosting stars discovered by the K2 mission via a quantitative high-resolution spectroscopic analysis based on a set of carefully vetted Fe I and Fe II lines.  As part of the analysis, we also derive chromospheric activity levels by using the Ca II H and K lines and investigate the impact that stellar activity might have on spectroscopically-derived stellar parameters. 

This paper is organized as follows: Section \ref{sec:data} provides details on the observations and data reduction. Section \ref{sec:st_par} discusses the methodology used to derive key stellar parameters and uncertainties, including effective temperatures, surface gravities, iron abundances (which can be taken as proxies for metallicities), stellar masses, radii, and planetary radii. In Section \ref{sec:S_ind} we explain the approach used to assess stellar activity using the calcium H and K lines and in Sections \ref{sec:discus} and \ref{sec:conclusion} present discussions and conclusions, respectively.

\section{Data} \label{sec:data}
The sample consists of host-stars observed by the Kepler extended mission (K2) during the C0 -- C8 campaigns. The optical spectra of these stars were obtained from the California Planet Search (CPS) program, which used the HIgh Resolution Echelle Spectrometer \citep[HIRES;][]{vogt1994} on the Keck I 10m telescope.  These spectra were observed during observational campaigns carried out after August 2004, with the updated HIRES CCDs, and cover the wavelength range from $\lambda$3640\AA\--7990\AA, which includes the Ca II H and K lines. 

Reduced HIRES spectra for a sample of 145 stars (one spectrum per star) were obtained from ExoFop\footnote{\url{https://exofop.ipac.caltech.edu/tess/}}\citep{exofop2}. These spectra were carefully inspected and 109 of them had signal-to-noise high enough in order to be analyzed for the derivation of stellar parameters and metallicities (Section \ref{sec:st_par}). As it was of interest to have as many observations as possible per star to study stellar activity, 725 spectra were obtained from the Keck Observatory Archives\footnote{\url{https://nexsci.caltech.edu/archives/koa/}} \citep[KOA;][]{koa1}, which were then used to measure the Ca II H and K lines located in the blue spectral region (Section \ref{sec:S_ind}).
Observational data, such as identifiers, positions, B and V magnitudes (taken from the EPIC catalog and the NASA Exoplanet Archive), and K2 observational campaigns information of the sample stars are given in Table \ref{tab:sample}.

\startlongtable
\begin{deluxetable*}{llccccc}
\tablecaption{Sample K2 Stars \label{tab:sample}}
\tablecolumns{7}
\tablenum{1}
\tablewidth{0pt}
\tablehead{
\colhead{ID} & \colhead{Host Name} & \colhead{R.A.} & \colhead{Decl.} & \colhead{$B$} & \colhead{$V$} & \colhead{Camp} \\
\colhead{} & \colhead{} & \colhead{(deg)} & \colhead{(deg)} & \colhead{(mag)} & \colhead{(mag)} & \colhead{}
}
\startdata
EPIC211319617 & K2-180 & 126.463933 & 10.246968 & 13.334 & 12.601 & C5 \\
EPIC211331236 & K2-117 & 133.855682 & 10.469131 & 16.137 & 14.655 & C5 \\
EPIC211342524 & \nodata & 128.098694 & 10.677239 & 13.052 & 12.422 & C5 \\
EPIC211351816 & K2-97 & 127.762836 & 10.847586 & 13.770 & 12.611 & C5 \\
EPIC211355342 & K2-181 & 127.554034 & 10.910294 & 13.494 & 12.749 & C5 \\
\nodata & \nodata & \nodata & \nodata & \nodata & \nodata & \nodata \\ 
\enddata
\tablecomments{This table is published in its entirety in the machine readable format. A portion is shown here for guidance regarding its form and content.}
\end{deluxetable*}

\section{Stellar Parameters} \label{sec:st_par}
The spectroscopic stellar parameters of effective temperature ($T_{\rm eff}$), surface gravity ($\log g$), iron abundance (A(Fe)\footnote{A(X)=log(N(X)/N(H))+12.0}), and microturbulent velocity ($\xi$), were derived by enforcing the excitation/ionization balances of a selected set of Fe I and Fe II equivalent widths. 
The analysis assumes local thermodynamic equilibrium (LTE) and uses 1-D plane parallel model atmospheres from the Kurucz ATLAS9 ODFNEW grid \citep{castelli2004a}.

The line list used in this study was adopted from \cite{yana2019MNRAS.490L..86Y}, which is based on the line list of \cite{Melendez2014}; it contains 61 Fe I and 13 Fe II lines. 
\cite{yana2019MNRAS.490L..86Y} provide a critical discussion of which Fe I and Fe II lines are most sensitive, along with those that are not sensitive, to magnetic activity through the combination of magnetic susceptibility, as measured by the Land\'e g-factor (g$_{L}$), and the line strength, as measured by the equivalent width.  Table \ref{tab:linelist} lists the wavelengths of the Fe I (species=26.0) and Fe II (species=26.1) lines along with their excitation potentials, $\log$ $gf$ values (both solar and from laboratory), and whether the line is classified as magnetically sensitive (Y) or not (N) by \cite{yana2019MNRAS.490L..86Y}. 
In the last column of Table \ref{tab:linelist} we show the equivalent widths (EWs) measured for the Solar spectrum (see discussion about the solar abundance determination below). We used the ARES code  v2 \citep{sousa2015A&A...577A..67S} to set the continuum and measure the EWs of all Fe lines in this study. Three line-free spectral regions: $\lambda$ 5764 – 5766 Å, $\lambda$ 6047 – 6052 Å, and $\lambda$ 6068 - 6076 Å were used to set the continuum level. In this section we only consider in the analysis Fe lines which are deemed as insensitive to stellar activity. (An analysis including sensitive lines will be discussed in Section \ref{sec:impact}).

The spectroscopic methodology adopted here is similar to that used in our previous study of K2 targets \citep{loaiza2023ApJ...946...61L}, which used an automated code for the determinations of stellar parameters and metallicities named \texttt{qoyllur-quipu}\footnote{\url{https://github.com/astroChasqui/q2}} or $q^2$ \citep{Ramirez2014}. Briefly, \textsc{$q^2$} uses an input iron line list and measured equivalent widths in combination with the 2019 version of the abundance analysis code MOOG \citep{sneden1973} to compute iron abundances, effective temperatures, surface gravities and metallicities.
The iterative process begins with an interpolated model atmosphere calculated for assumed values of $T_{\rm eff}$, $\log g$, and metallicity.  Values of $T_{\rm eff}$ / $\log g$ / A(Fe) are adjusted iteratively (increased or decreased) to minimize trends A(Fe I) and A(Fe II) as functions of $T _{\rm eff}$, $\log g$, and $\xi$ until a solution is found for a final adjusted set of spectroscopic parameters of each star.

As a test to our methodology we analyzed solar-proxy spectra obtained with the HIRES spectrograph, focusing on reflected solar light from the asteroid Iris. The parameters and metallicities obtained for the solar proxy Iris were: $T_{\rm eff}= 5744\pm33$ K, $\log g = 4.52\pm0.05$, A(Fe) $= 7.39\pm0.02$, and $\xi=1.12 \pm 0.13$ km.s$^{-1}$; these are in general agreement with the solar parameters, indicating that our methodology does not likely have strong biases for solar-type stars. We note, however, that the mean metallicity for the solar proxies of A(Fe) = 7.39 is slightly more metal-poor than the \cite{asplund2021AA...653A.141A} (A(Fe)$_\odot = 7.46$) scale, keeping in mind that they obtain A(Fe) $<$ 7.40 when they analyze in 1D.

\begin{figure}
\centering
\epsscale{1.2}
\plotone{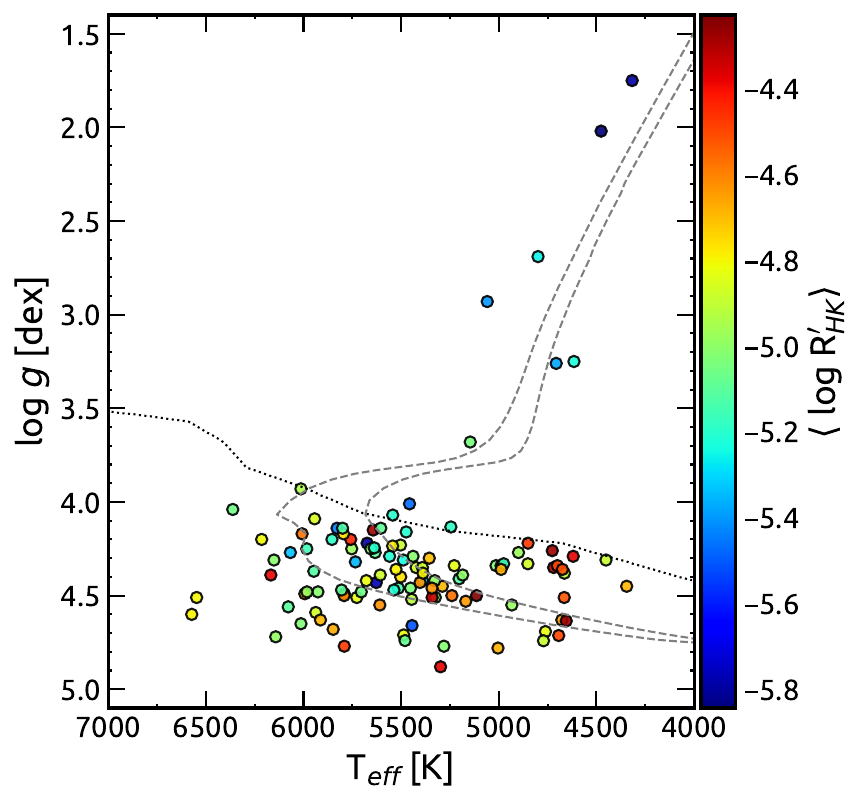}
\caption{Kiel diagram for the sample stars. The gray dashed lines represent solar metallicity 4.6 and 10 Gyr old stellar evolution tracks from Yonsei-Yale. The colors of the circles indicate stellar activity as measured by the $\log R^\prime_{\rm HK}$ indicator (Section \ref{sec:S_ind}) and corresponding to the color bar. The main sequence stars and the evolved stars are separated by the black dashed line.} 
\label{fig:kiel_sample}
\end{figure}

The derived effective temperatures, surface gravities, metallicities, and microturbulent velocities for all analyzed stars are presented in Table \ref{tab:parameters}.
We show the stellar parameters of $T_{\rm eff}$ and $\log g$ for the K2 stars in this study in the Kiel diagram presented in Figure \ref{fig:kiel_sample}. The target stars are shown as filled circles color-coded by their $\log R^\prime_{\rm HK}$ index (which will be derived in Section \ref{sec:S_ind}) as indicated by the color bar. We also show two Yonsei-Yale isochrones \citep{yi2001,yi2003ApJS..144..259Y,demarque2004ApJS..155..667D,han2009gcgg.book...33H} corresponding to 4.6 and 10 Gyr as dashed gray lines. The majority of the stars in this study are on the main sequence, with a small number of them having lower values of $\log g$ which are indicative of evolved stars.
The black dashed line in Figure \ref{fig:kiel_sample} represents the adopted division between main-sequence and evolved stars. Here we use the same division between main sequence and subgiants and giants that was adopted in \cite{gomes2021AA...646A..77G} which is from \cite{ciardi2011AJ....141..108C}.

Comparisons of the derived stellar parameters with results from the literature find, in general, reasonable agreement. There is good agreement between our values of $T_{\rm eff}$s with those from the works of \cite{petigura2018} and \cite{mayo2018}, as well as the DR17 APOGEE \citep{majewski2017} and the DR3 GALAH \citep{Buder2021} surveys. 
Comparisons of the different temperature scales find median differences values of $T_{\rm eff}$ (``Other Work" - ``This Work") and MADs (Median Absolute Deviations) of: $+22 \pm 71$ and $+35 \pm 74$ K for the two first studies, and $-16 \pm 66$ and $-52 \pm 65$ K for the two surveys, respectively. The comparison with $T_{\rm eff}$s from the Gaia survey \citep{gaia2016AA...595A...1G,gaia2023AA...674A...1G} and the DR8 LAMOST survey \citep{cui2012} reveals slightly more significant systematic differences and MADs of $-52 \pm 95$ and $-62 \pm 107$ K, respectively. 

Our values of $\log g$ also generally agree, within the uncertainties, with those from the mentioned works/surveys, with median differences in $\log g$ from $\sim -$0.02 to 0.02 dex and MAD values from $\pm$ 0.13 to $\pm$ 0.18.
For \cite{mayo2018} and LAMOST, however, we find higher systematic differences of +0.16 $\pm$ 0.12 and $-$0.03 $\pm$ 0.14 dex, respectively. 
Finally, our $\log g$ values overall compare well with those based on asteroseismology for K2 stars from \cite{huber2016ApJS..224....2H}: median $\log g$ difference ($\pm$ MAD) of $-$0.01 $\pm$ 0.22 dex. But we note that there are four stars with significant differences, with this work finding them to be dwarfs (log g varying between 4.3 to 4.8 with uncertainties between 0.08 and 0.2), while \cite{huber2016ApJS..224....2H} find them to be more evolved ($\log g$ varying between 2.7 to 3.7 with uncertainties from 0.07 to 0.53 dex). These are: EPIC 220481411, EPIC 211413752, EPIC 212012119, and EPIC 212782836, which are nearby stars with distances of approximately 114, 321, 123, and 184 pc, and absolute $V$-magnitudes of 7.04, 6.39, 6.69, and 5.79, respectively, placing them as K-dwarf stars and more in line with our $\log g$ values. (See also discussion in Section \ref{sec:radii} of differences in the derived radii with \cite{huber2016ApJS..224....2H}).
\vspace{-\baselineskip}
\begin{deluxetable}{lccccccccccccc}
\tiny
\tablecaption{Iron Line List\label{tab:linelist}}
\tablecolumns{12}
\tablenum{2}
\tablewidth{0pt}
\tablehead{
\colhead{$\lambda$} & \colhead{Species} & \colhead{$\chi$} & \colhead{$\log$ $gf$} & \colhead{$Sensitive$} & \colhead{EW$_\odot$}\\
\colhead{(\AA)} & \colhead{} & \colhead{(eV)} & \colhead{} & \colhead{} & \colhead{(m\AA)} & \colhead{} & \colhead{}
}
\startdata
4088.560	&	26.0	&	3.640	&	-1.720	&	Y	& \nodata \\
4091.560	&	26.0	&	2.830	&	-2.310	&	Y	& \nodata \\
4365.896	&	26.0	&	2.990	&	-2.250	&	N	& 50.83 \\
4389.245	&	26.0	&	0.052	&	-4.583	&	Y	& \nodata \\
4445.471	&	26.0	&	0.087	&	-5.441	&	N	& 41.07 \\
4485.970	&	26.0	&	3.640	&	-2.530	&	Y	& \nodata \\
\nodata	&	\nodata	&	\nodata	&	\nodata	&	\nodata \\
\enddata
\tablecomments{The complete table is available in machine-readable format. Here, a portion is provided for reference regarding its format and content. Lines with (Y) are sensitive to activity. Lines with (N) are not sensitive.
Lines marked with (*) were excluded from the analysis of stellar parameters for all stars because their equivalent widths exceeded 120 m\AA.}
\end{deluxetable}

{\small
\begin{deluxetable*}{lcccccccccccc}
\tablecaption{Stellar Parameters \label{tab:parameters}}
\tablecolumns{13}
\tablenum{3}
\tablewidth{0pt}
\tablehead{
\colhead{ID} & \colhead{$\langle\rm S_{HK}\rangle$} & \colhead{$\sigma$} & \colhead{$\langle\rm log R^\prime_{HK}\rangle$} & \colhead{$\sigma$} & \colhead{\# Sp} & \colhead{$T_{\rm eff}$} & \colhead{log g} & \colhead{A(Fe)} & \colhead{$\xi$} & \colhead{$\rm R_{star}$} & \colhead{$\rm M_{star}$} & \colhead{$v \sin i$} \\
\colhead{} & \colhead{} & \colhead{} & \colhead{} & \colhead{} & \colhead{} & \colhead{(K)} & \colhead{(dex)} & \colhead{(dex)} & \colhead{($\rm km.s^{-1})$} & \colhead{($\rm R_\odot$)} & \colhead{($\rm M_\odot$)} & \colhead{($\rm km.s^{-1})$}
}
\startdata
EPIC211319617 &  0.166 &  0.019 & -5.024 &  0.123 &   15 &   5281 $\pm$  54 & 4.77 $\pm$ 0.15 & 6.82 $\pm$ 0.03 & 0.89 $\pm$ 0.29 & 0.674 $\pm$ 0.015 & 0.711 $\pm$ 0.018 & 1.64 (1) \\
EPIC211331236 &  2.488 &  \nodata & -4.648 &  \nodata &    1 &   \nodata & \nodata & \nodata & \nodata & \nodata $\pm$   NaN & \nodata $\pm$ \nodata & 2.91 (1) \\
EPIC211342524 &  0.259 &  \nodata & -4.571 &  \nodata &    1 &   6007 $\pm$ 163 & 4.17 $\pm$ 0.28 & 7.08 $\pm$ 0.11 & 2.53 $\pm$ 1.64 & 1.611 $\pm$ 0.097 & 1.031 $\pm$ 0.047 & 7.80 (2) \\
EPIC211351816 &  0.110 &  0.023 & -5.374 &  0.108 &    7 &   4706 $\pm$  74 & 3.26 $\pm$ 0.22 & 7.62 $\pm$ 0.07 & 1.44 $\pm$ 0.14 & 3.699 $\pm$ 0.132 & 1.118 $\pm$ 0.079 & 3.70 (2) \\
EPIC211355342 &  0.149 &  0.017 & -5.139 &  0.138 &    9 &   5514 $\pm$  23 & 4.46 $\pm$ 0.06 & 7.62 $\pm$ 0.03 & 1.02 $\pm$ 0.14 & 1.012 $\pm$ 0.020 & 0.953 $\pm$ 0.017 & 1.63 (1) \\
\nodata & \nodata & \nodata & \nodata & \nodata & \nodata & \nodata & \nodata & \nodata & \nodata & \nodata & \nodata & \nodata \\
\enddata
\tablecomments{The $v \sin i$ values come from the literature, APOGEE DR17 (1) and \cite{petigura2018} (2). This table is published in its entirety in the machine readable format. A portion is shown here for guidance regarding its form and content.}
\end{deluxetable*}}

\vspace{-\baselineskip}
\subsection{Stellar Masses \& Radii} \label{sec:radii}

We calculated the stellar masses and radii using the isochrone method, which involves estimating these fundamental parameters by comparing the positions of stars in a $T_{\rm eff}$ -- $M_V$ diagram with theoretical isochrones. To perform this analysis, we used the PARAM code v1.3 \citep{dasilva2006AA...458..609D} 
which derives stellar mass, radius, luminosity, age, etc., based on a grid of PARSEC isochrones \citep{bressan2012}   
employing Bayesian inference methodology.
The required input parameters are $T_{\rm eff}$ and [Fe/H], determined spectroscopically in this work, and the absolute magnitude ($M_V$) derived using the parallax measurements obtained from Gaia DR3 data \citep{gaia2021A&A...649A...1G}.  
In Table \ref{tab:parameters} are shown the derived masses and radii of the sample stars.

As a comparison, we also determined the stellar radii and masses for the studied stars using the code q$^2$ (as in \citet{loaiza2023ApJ...946...61L}), which is based on a grid of Yonsei-Yale isochrones \citep{yi2001,yi2003ApJS..144..259Y}. The results obtained with q$^2$ are in good agreement with those from PARAM. The median differences ``PARAM -- q$^2$'' are $-$0.03 $\pm$ 0.02 R$_\odot$ for radii and $-$0.01 $\pm$ 0.01 M$_\odot$ for masses. 

\begin{figure*}[!]
\epsscale{1}
\plotone{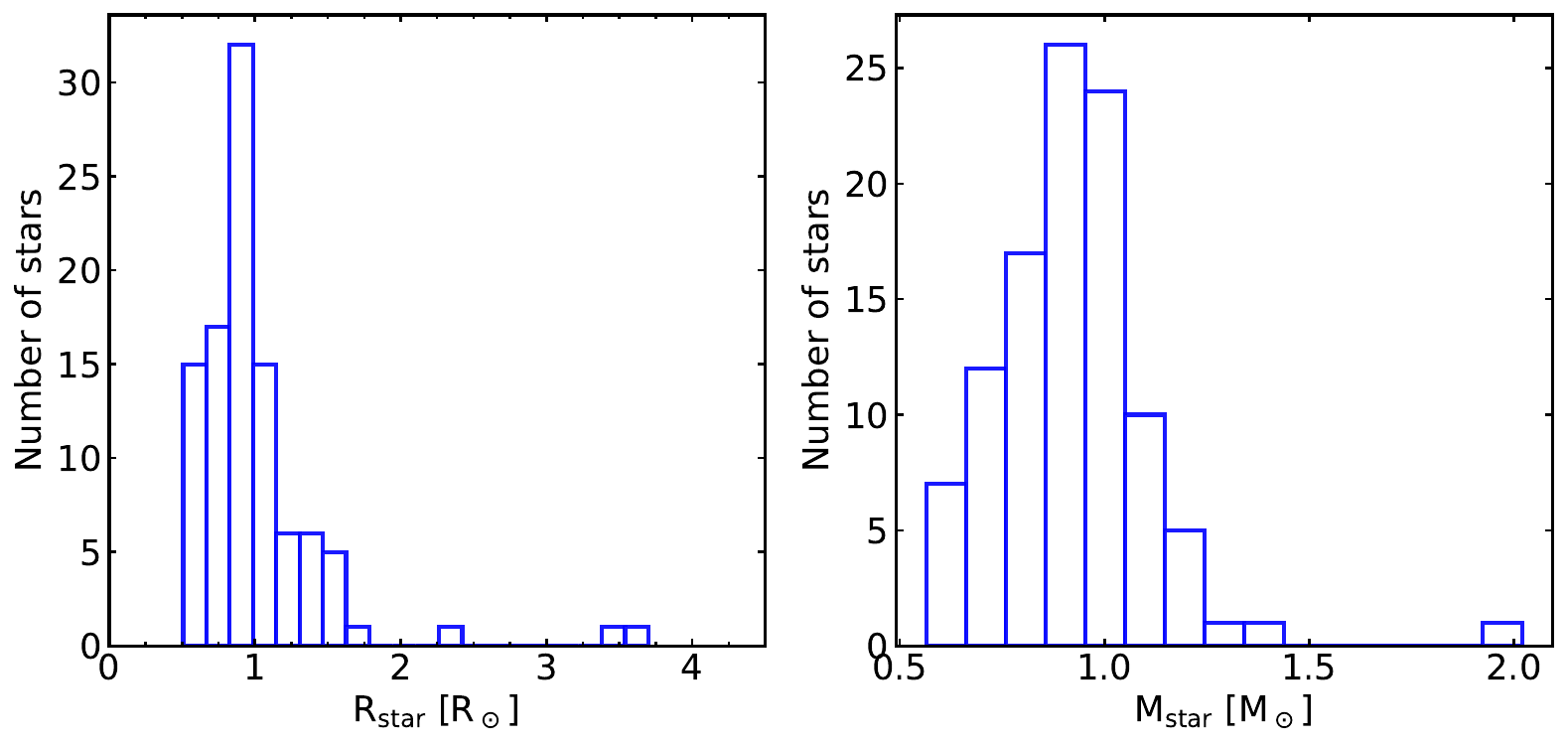}
\caption{Mass and radius distributions for the sample stars. For better visualization, those few stars in our sample with stellar radii larger than 4 R$_\odot$ are not shown in the figure.}
\label{fig:RM_dist}
\end{figure*}

The distributions of stellar masses and radii for our sample are displayed in Figure \ref{fig:RM_dist}, revealing that the majority of the sample stars have stellar radii $<2$ $\rm R_\odot$ and masses $< 1.2$ $\rm M_\odot$. The median radius of the distribution is 0.92 R$_\odot$ (16th percentile = 0.70; 84th percentile = 1.35) and median mass is 0.92 M$_\odot$ (16th percentile = 0.74; 84th percentile = 1.05).

\begin{figure*}
\gridline{\fig{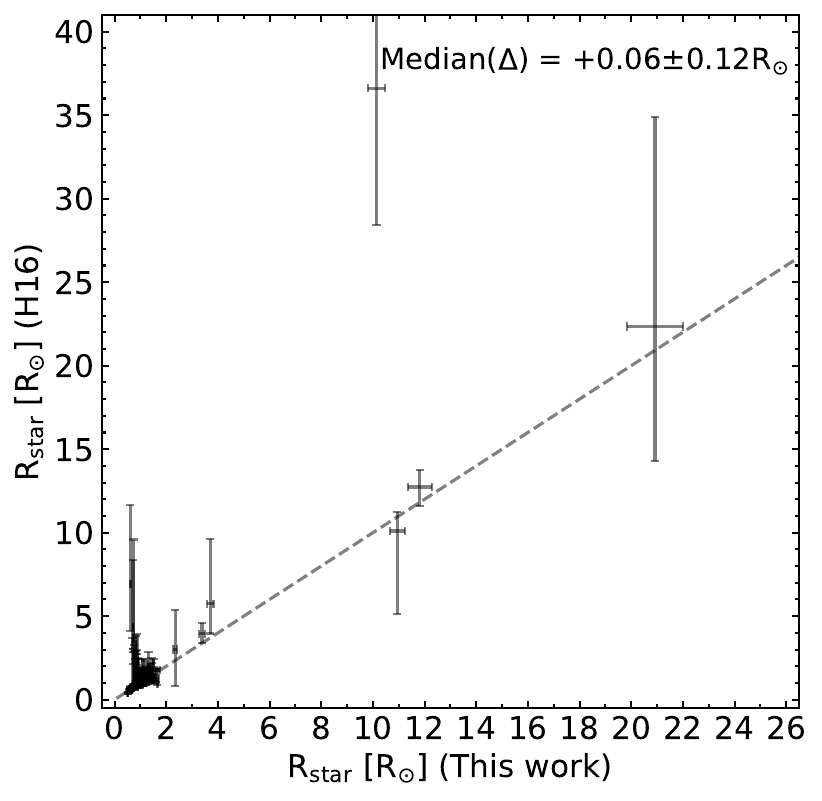}{0.49\textwidth}{}
            \hspace{-3mm}
          \fig{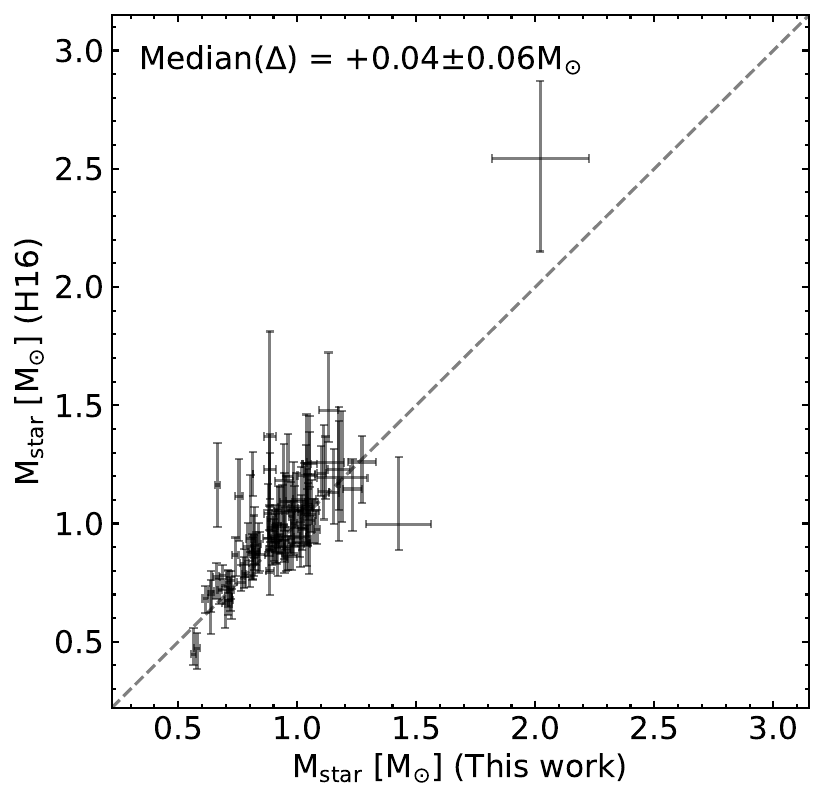}{0.49\textwidth}{}
           }
        \vspace{-8mm}
\caption{Comparisons of the stellar radii and masses in this study with those determined through asteroseismology by \citet[][H16]{huber2016ApJS..224....2H}. The median differences (`Other Work - This Work') between the parameters and the corresponding MAD are indicated in each case.}\label{fig:Rstar_Mstar_comp}
\end{figure*}

The stellar masses and radii obtained here using the isochrone method are compared with results from asteroseismology from \cite{huber2016ApJS..224....2H} in Figure \ref{fig:Rstar_Mstar_comp}, with stellar radii shown on the left panel and masses on the right panel.

Overall the derived radii and masses compare well with \citet{huber2016ApJS..224....2H}, with a small systematic difference, except for a few clear outliers. The median ($\pm$ MAD) of the differences between the stellar radii from \cite{huber2016ApJS..224....2H} minus ours is 0.06 $\pm$ 0.12 R$_\odot$.  
We note, however, that there seems to be a slight trend for the more evolved stars having radii roughly larger than $\sim$4 R$_\odot$. Concerning the masses, a small systematic difference is found between our stellar mass determinations and those of \cite{huber2016ApJS..224....2H} (see right panel Figure \ref{fig:Rstar_Mstar_comp}). Our masses are smaller than those from asteroseismology with a small median ($\pm$ MAD) of the differences of 0.04 $\pm$ 0.06 M$_\odot$.

One of the outliers in the radius comparison with \citet{huber2016ApJS..224....2H} is particularly significant - star EPIC 215346008, for which we obtained a stellar radius of 10.141 $\pm$ 0.342 R$_\odot$ which is about 3 times smaller than the one in \citet{huber2016ApJS..224....2H} of 36.613 R$_\odot$, but the latter has an extremely large uncertainty of +86.878 and $-$8.192 R$_\odot$.  
This star is also reported in \cite{petigura2018} as having 12.34 R$_\odot$, which is more similar to ours, while in \cite{Hardegree2020}, who obtained radius from Stefan-Boltzmann law using stellar parameters from the LAMOST survey, this star has 24.224 R$_\odot$. Finally, we note that in Gaia DR3 the radius for this star is 29.53 R$_\odot$ with a 68\% confidence interval ranging from 28.29 R$_\odot$ to 30.52 R$_\odot$, based on the 16th and 84th percentiles, respectively.  
Another star with a large difference in radius in this comparison is EPIC 220481411, for which we obtain 0.609 $\pm$ 0.012 R$_\odot$,   
while \cite{huber2016ApJS..224....2H} obtain 6.924$^{+4.716}_{-2.805}$ R$_\odot$. 
Other results in the literature also find radii in better agreement with our result: \citet[][0.69 R$_\odot$]{livingston2018}, \citet[][ 0.72 R$_\odot$]{persson2018AA...618A..33P}, \citet[][ 0.67 R$_\odot$]{mayo2018}, \citet[][0.71 R$_\odot$]{petigura2018}, and \citet[][0.64 R$_\odot$]{kruse2019}.
EPIC 220481411 is a nearby star with a well-determined parallax from Gaia DR3 of $\pi$=8.69 mas, with a distance of $\sim$115 pc 
and, with an apparent $V$-magnitude of 12.48, the absolute $V$-magnitude is $M_{V}$=7.15, placing this star squarely in the realm of the K-dwarfs, where a sub-solar radius is expected, giving confidence to our derived radius.

Concerning the mass comparison, the most significant outlier in Figure \ref{fig:Rstar_Mstar_comp} is the star EPIC 220503133, for which we find 2.021 $\pm$ 0.203 M$_\odot$,  
while the asteroseismic result from \cite{huber2016ApJS..224....2H} is 2.543$^{+0.328} _{-0.393}$ M$_\odot$, with results overlapping within the error bars.

\subsection{Planetary Radii} \label{sec:planet}
The planetary radii were determined using the inferred stellar radii and the transit depth ($\Delta F$), which represents the fraction of stellar flux lost during the planet's transit minimum, given by \cite{seager2003}:

\begin{equation}
    R_{pl} = 109.1979 \times \sqrt{\Delta F \times 10^{-6}} \times R_{star}, 
\end{equation}
where the planet's radius is given in terms of Earth radii.

\begin{figure*}
\gridline{\fig{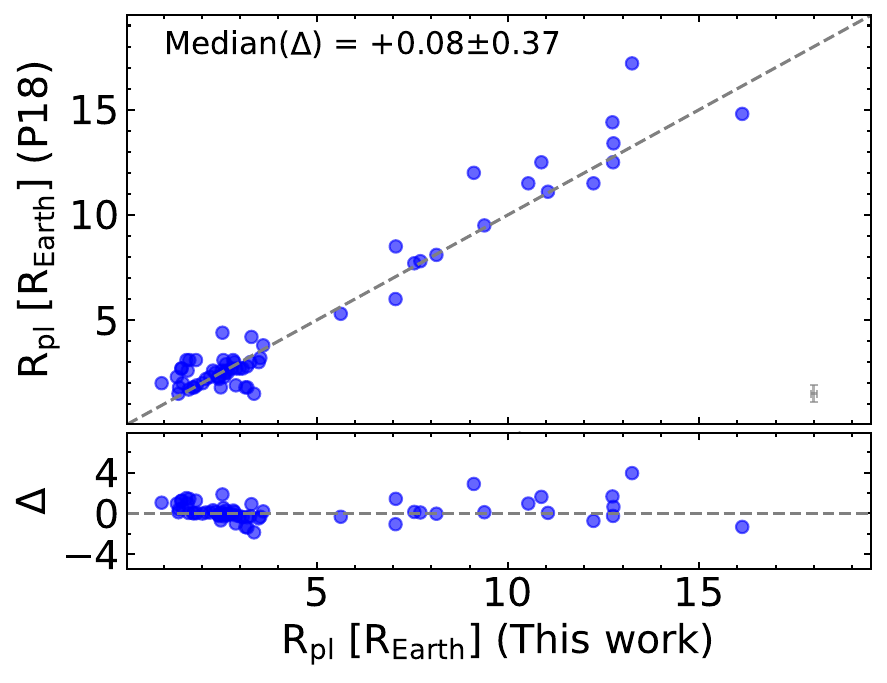}{0.47\textwidth}{}
            \hspace{-3mm}
          \fig{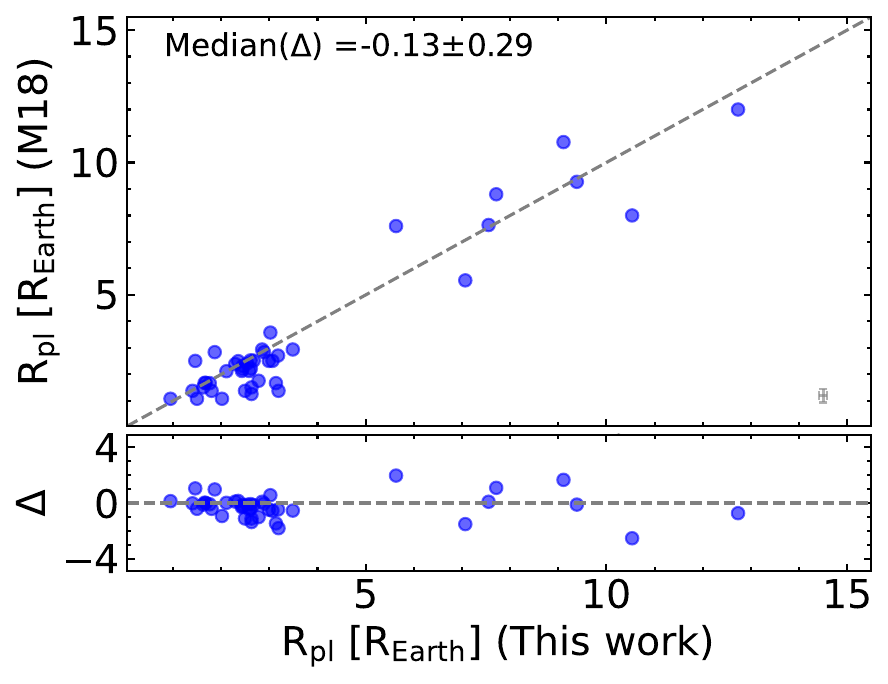}{0.47\textwidth}{}
           }
        \vspace{-8mm}
\gridline{\fig{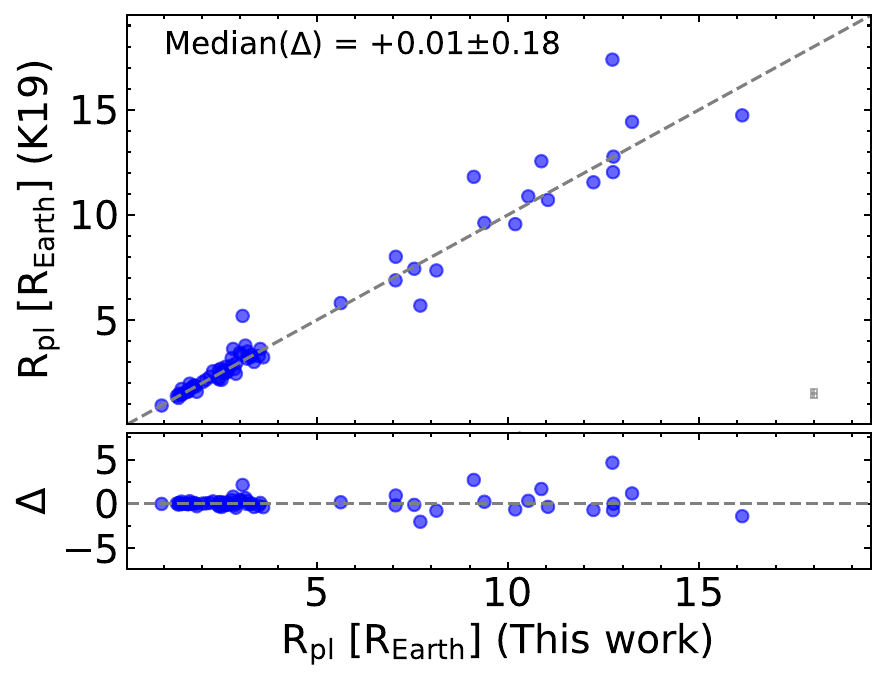}{0.47\textwidth}{}
            \hspace{-3mm}
          \fig{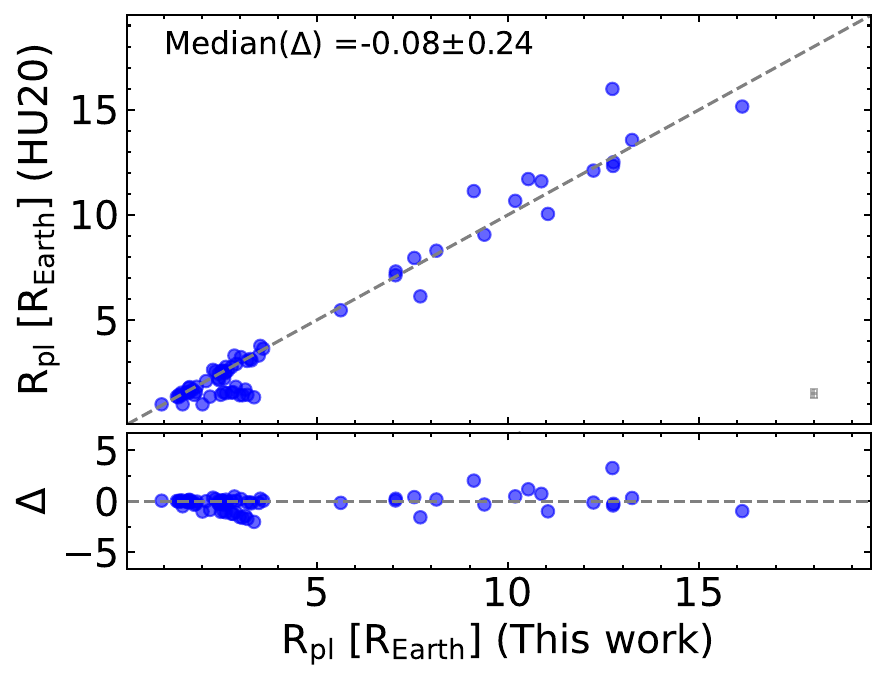}{0.47\textwidth}{}
           }
          \vspace{-8.5mm}
\gridline{\fig{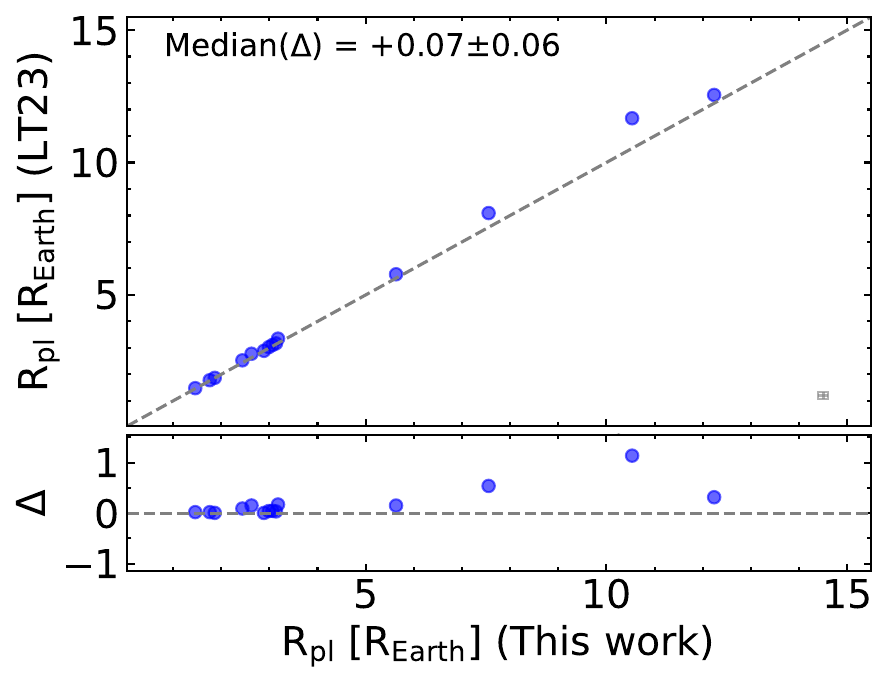}{0.47\textwidth}{}
           }
        \vspace{-8mm}
\caption{Comparisons of the planetary radii in this study with those from  \citet[][P18,]{petigura2018}, \citet[][M18]{mayo2018}, \citet[][K19]{kruse2019}, \citet[][HU20]{Hardegree2020}, and \citet[][LT23]{loaiza2023ApJ...946...61L}. The bottom sub-panels show the difference between `Other Work - This Work' ($\Delta$). The median differences between the parameters and the corresponding MAD are indicated in each case. The black dashed lines represent equality.}
\label{fig:Rpl_comp}
\end{figure*}

For our analysis, we considered only the $\Delta F$ values of confirmed planets. All candidates or false positives, according to the NASA Exoplanet Archive notes, were excluded from our planet sample. 
The $\Delta F$ values were obtained from \cite{kruse2019}, \cite{barros2016AA...594A.100B}, \cite{pope2016MNRAS.461.3399P}, and \cite{christiansen2017AJ....154..122C}, and these values, along with the star identifications, planet names, planetary radii ($\rm R_{pl}$) and their corresponding errors ($\rm \delta R_{pl}$), are listed in Table \ref{tab:planet_rad}. 
The comparison between the radii of exoplanets orbiting K2 host stars determined here and those in \cite{loaiza2023ApJ...946...61L}, \cite{Hardegree2020}, \cite{kruse2019}, \cite{petigura2018}, and \cite{mayo2018} is shown in Figure \ref{fig:Rpl_comp}. 
In general, the median differences in planetary radii (``Other Work -- This Work'') are small, varying between $\sim$ $-$0.13 and $\sim$ +0.08; more specifically corresponding to 6\% difference for \cite{kruse2019}, 7.6\% for \cite{Hardegree2020}, 10.7\% for \citep{petigura2018}, 11.1\% for \cite{mayo2018}, and 1.3\% for \cite{loaiza2023ApJ...946...61L}.

\begin{deluxetable}{llccc}
\tablecaption{Planetary Radii\label{tab:planet_rad}}
\tablecolumns{11}
\tablenum{4}
\tablewidth{0pt}
\tablehead{
\colhead{Star ID} & \colhead{Planet Name} & $\Delta F$ &   \colhead{$R_{pl}$} & \colhead{$\delta R_{pl}$} \\
\colhead{} & \colhead{}  &  (ppm) & \colhead{($R_\oplus$)} & \colhead{($R_\oplus$)}
}
\startdata
EPIC211319617  &  K2-180 b         &     1189.0 &     2.54 &     0.07 \\
EPIC211351816  &  K2-97 b          &      680.0 &    10.53 &     0.44 \\
EPIC211355342  &  K2-181 b         &      828.0 &     3.18 &     0.08 \\
EPIC211399359  &  K2-371 b         &    28460.0 &    12.75 &     0.03 \\
EPIC211413752  &  K2-268 b         &      398.0 &     1.60 &     0.08 \\
EPIC211413752  &  K2-268 c         &     1239.0 &     2.82 &     0.08 \\
EPIC211413752  &  K2-268 d         &      531.0 &     1.84 &     0.06 \\
EPIC211413752  &  K2-268 e         &      436.0 &     1.67 &     0.06 \\
EPIC211413752  &  K2-268 f         &     1025.0 &     2.56 &     0.09 \\
\nodata  & \nodata  & \nodata  & \nodata     \\    
\enddata
\tablecomments{Transit depth ($\Delta F$) collected from \cite{kruse2019}, \cite{pope2016MNRAS.461.3399P}, \cite{barros2016AA...594A.100B}, and \cite{christiansen2017AJ....154..122C} for K2 planets. This table is published in its entirety in the machine readable format. A portion is shown here for guidance regarding its form and content.}
\end{deluxetable}

\subsubsection{Planetary Radii and the Radius Gap} \label{sec:pl_radii}
\begin{figure}
    \centering
    \includegraphics[scale=0.55]{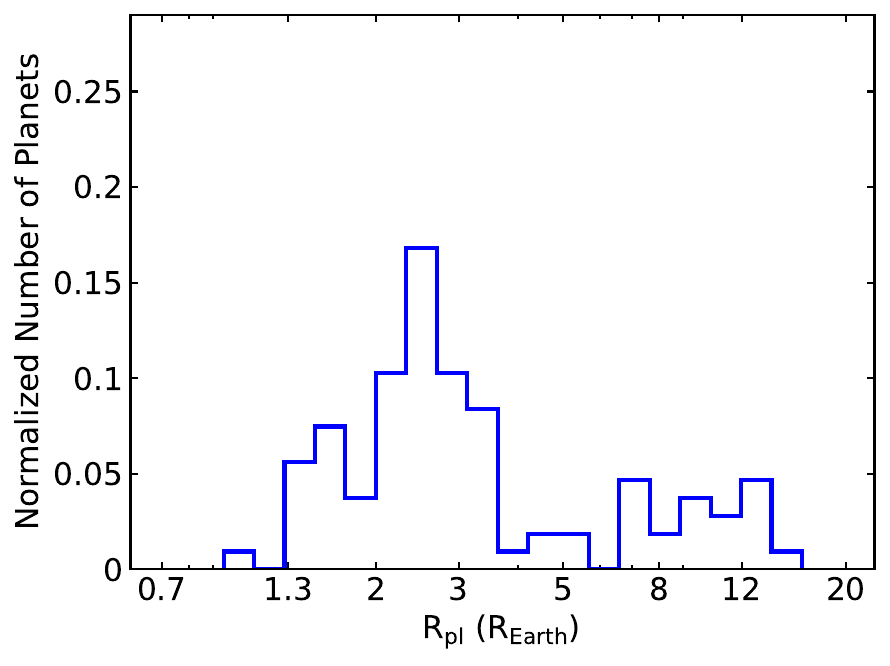}
    \caption{The planetary radius distribution for the K2 planet sample. The location of the gap in the planetary radius distributions is approximately at R$\rm _{gap} \sim 1.9$ $R_\oplus$.} 
    \label{fig:Rpl_gap}
\end{figure}

Figure \ref{fig:Rpl_gap} shows the distribution of planetary radii for our sample of 93 confirmed planets orbiting 69 K2 stars. Not unexpectedly, there is a dearth of planets around $\rm R_{gap} \sim 1.9$ $R_\oplus$ for the smaller planets in the sample (R$_{pl} < 4$ R$_\oplus$). 
Similarly to what was found in our previous study of K2 targets based on  Hydra spectra \citep{loaiza2023ApJ...946...61L} and with a similar number of confirmed planets, our methodology for determining stellar radii and, from them, planetary radii allowed for the identification of the radius gap, also known as the Fulton gap \citep{fulton2017}. This well known feature has been widely confirmed in various studies using the Kepler sample \citep{berger2018,fulton2018,vaneylen2018,Martinez2019} and the fact that our results depict the gap attests to the high quality and precision achieved by our spectroscopic analysis. The error budget in the derived radii are discussed in the section below.

\subsection{Uncertainties in the Derived Parameters} \label{sec:errors}
The formal errors adopted for the stellar parameters $T_{\rm eff}$, $\log g$, and $\xi$ were computed using $q^2$, which follows the error analysis discussed in \cite{epstein2010} and \cite{bensby2014}.
Errors in the iron abundances, A(Fe I) and A(Fe II), were obtained by combining errors estimated from the equivalent-width measurements with stellar parameter uncertainties. 
The individual errors of these parameters are presented in Table \ref{tab:parameters}.

The median errors in the stellar parameters derived in this study are reported in Table \ref{tab:error_budget}: $\delta T_{\rm eff}=42$ K, $\delta \log g=0.09$ dex, $\rm \delta A(Fe) = 0.03$ dex, and $\rm \delta \xi = 0.17$ $\rm km.s^{-1}$. Following a similar approach to that employed by \cite{loaiza2023ApJ...946...61L}, we have taken into account the contributions of individual errors in determining the error budget for derived parameters, such as $\rm M_{star}$, $\rm R_{star}$, and $\rm R_{pl}$. It is important to note that the error in the $V$ magnitude contributes about 0.15\% to the error in the stellar radius, assuming a median error of 0.04 mag for the $V$ magnitudes of our stars. In addition, errors in parallaxes, with a median error of 0.02 mas, result in a 0.66\% error in both stellar mass and radius. The uncertainties in stellar radii and transit depth errors ($\rm \Delta F$) have a direct influence on the determination of planetary radius errors. The median internal uncertainty in our derived stellar radius distribution is 1.8\%. For the transit depth values ($\rm \Delta F$) and their associated errors, we adopted data from \cite{kruse2019}, \cite{barros2016AA...594A.100B}, and \cite{christiansen2017AJ....154..122C}, resulting in an internal transit depth precision of 2.8\% for the planets in our sample. Ultimately, these combined uncertainties contribute to an internal precision of 2.3\% in the error budget for the radius of the planets ($\rm R_{pl}$). Table \ref{tab:error_budget} summarizes the error budgets for our $\rm R_{star}$ and $\rm R_{pl}$ determinations.

\begin{deluxetable}{ll}
\tablecaption{Error budget\label{tab:error_budget}}
\tablecolumns{2}
\tablenum{5}
\tablewidth{0pt}
\tablehead{
\colhead{Parameter} & \colhead{Median Uncertainty}
}
\startdata
$T_{\rm eff}$  & 42 K \\
$\log g$ & 0.09 dex \\
A(Fe)         & 0.03 dex \\
$V$             & 0.04 mag \\
plx           & 0.02 mas \\
M$_{star}$    & 0.02  M$_\odot$ \\
R$_{star}$     & 1.79\% \\
$\Delta$F     & 2.79\% \\
R$_{pl}$      & 2.33\% \\
\enddata
\end{deluxetable}

\section{Measurements of the Calcium H and K Lines} \label{sec:S_ind}
The Ca II H and K lines are useful diagnostics of chromospheric activity due to the sensitivity of their line cores to chromospheric temperature inversions, which are related to magnetic activity \citep{leighton1959ApJ...130..366L}.  Quantitative Ca II H and K activity levels can be expressed using the S-index, also known as the Mount Wilson index \citep[][S$_{MW}$]{Vaughan1978PASP...90..267V}. 
The S-index is defined as a ratio of fluxes measured in the H and K line cores and fluxes in nearby pseudo-continuum passbands, which are labeled R and V. 
In this study these fluxes were measured following the procedure described in \cite{isaacson2010ApJ...725..875I} and using the IRAF\footnote{Image Reduction and Analysis Facility (IRAF) is distributed by the National Optical Astronomy Observatory, which is operated by the Association of Universities for Research in Astronomy, Inc., under a cooperative agreement with the National Science Foundation.} task \texttt{sbands}.  Fluxes in the H and K line cores were measured using a triangular bandpasses (full width at half maximum of 1.09 \AA) centered at $\lambda$3968.47 \AA\ and $\lambda$3933.67 \AA, respectively, while the pseudo-continuum R and V fluxes were measured using rectangular bandpasses of 20 \AA, centered at $\lambda$3901 \AA\ and $\lambda$4001 \AA, respectively.

\begin{figure*}
    \centering
   \includegraphics[scale=0.32]{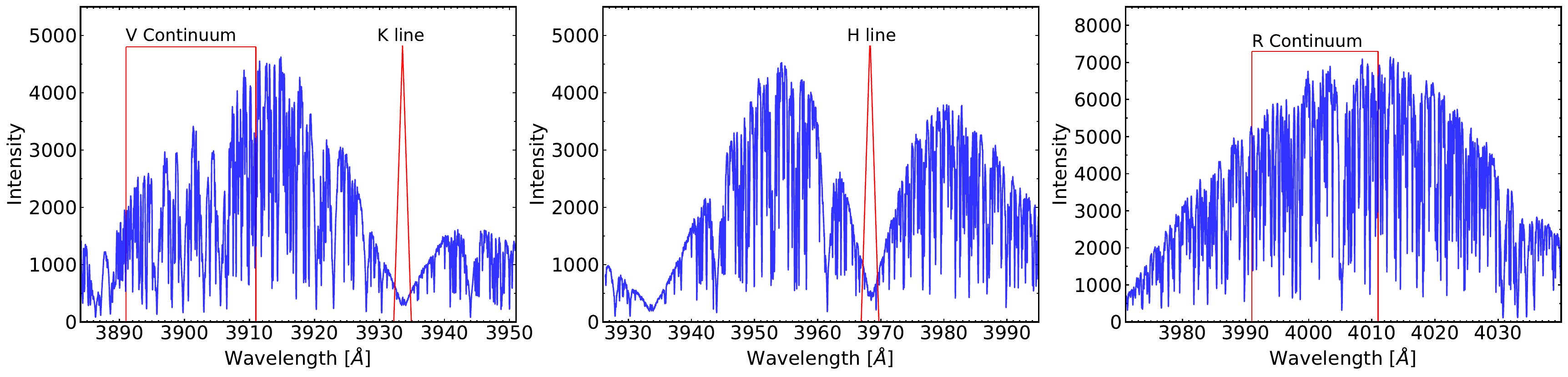}
   \caption{Keck HIRES spectra of the star EPIC220383386 showing the three echelle orders containing the calcium H and K lines and pseudo-continuum at V and R.}
   \label{fig:ca_hk}
\end{figure*}

\cite{isaacson2010ApJ...725..875I} provide a defining equation for the S$_{\rm HK}$-index from HIRES spectra of

\begin{equation}\label{eq:shk}
    \rm S_{HK} = 32.510 \frac{(H +1.45K)}{(R +25V)} + 0.021
\end{equation}

where the coefficient for V was set in order to bring the mean fluxes in the R and V continuum bandpasses to approximately the same levels. Our use of this equation led to unrealistically small values of S$_{\rm HK}$ for our HIRES spectra.  Following the discussion in \cite{isaacson2010ApJ...725..875I}, we investigated what V-coefficient would bring the measured R and V mean-fluxes to approximately the same levels in our HIRES spectra (see Figure \ref{fig:ca_hk}), and found a mean V-coefficient from all spectra of 2.28, with a small scatter.  The computation of S$_{\rm HK}$ used a unique V-coefficient for each spectrum such that the R and V mean fluxes were equal.

We measured the $\rm S_{HK}$ index in a total of 725 spectra from 144 K2 stars, noting that for 40 of our targets there were two or more spectra available in the Keck Archive, with the other stars having only one HIRES observation. The measured $\rm S_{HK}$ index for the stars are presented in Table \ref{tab:parameters}, noting that for those stars having more than one observation, the values in the table are the mean (and standard deviation) of the S$\rm _{HK}$ measurements. The $\rm S_{HK}$ indices for our K2 sample will the discussed in Section \ref{sec:discus} and validated from comparisons with the literature below.

\begin{figure}
    \centering
    \includegraphics[scale=0.65]{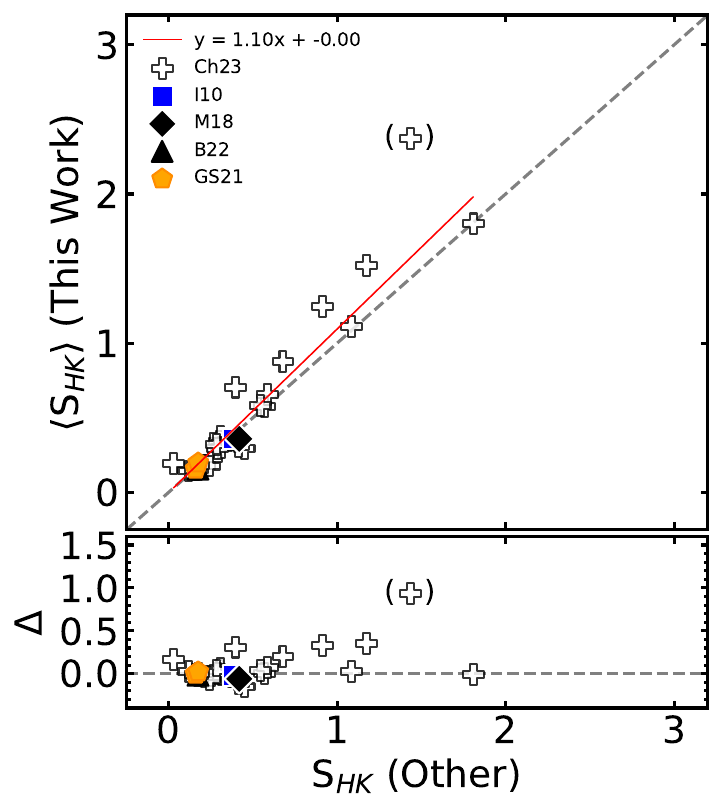}
    \caption{Comparison of the derived S-indices for stars in common with other studies in the literature. The crosses correspond to \citet[][Ch23]{chahal2023MNRAS.525.4026C}, open circles to \citet[][Z22]{zhang2022ApJS..263...12Z}, triangles to \citet[][B22]{brown2022MNRAS.514.4300B}, pentagon to \citet[][GS21]{gomes2021AA...646A..77G}, diamond to \citet[][M18]{mayo2018}, and  square to \citet[][I10]{isaacson2010ApJ...725..875I}. The gray dashed line in the upper panel represents equality, while the red solid line represents the best fit between our results and the literature.}
    \label{fig:SHK_comp}
\end{figure}

We opted to place our $\rm S_{HK}$ measurements from archival spectra onto the commonly used Mt. Wilson scale via a comparison for stars in common with the following studies, all of which have transformed their respective $\rm S_{HK}$ values onto the Mt. Wilson scale: \cite{chahal2023MNRAS.525.4026C}, who used LAMOST spectra, \cite{brown2022MNRAS.514.4300B}, who used spectra from NARVAL and ESPaDOnS, \cite{gomes2021AA...646A..77G}, \cite{mayo2018}, and \cite{isaacson2010ApJ...725..875I} who used spectra from HARPS, TRES, and HIRES, respectively. 
The comparisons are shown in Figure \ref{fig:SHK_comp}, with the largest number of stars in common being \cite{chahal2023MNRAS.525.4026C}, where the median of the differences (and MAD) between $\rm S_{HK}$(``Chahal - This Work'') are 
$-0.033 \pm 0.052$, respectively. These are small differences and the overall comparisons with \cite{chahal2023MNRAS.525.4026C}, as well as for the other studies, are good.  The solid red line in Figure \ref{fig:SHK_comp} is the least squares fit of $\rm S_{HK}$(This Work) as a function of $\rm S_{HK}$(Other), while the dashed line is a one-to-one line.  The fitted slope of m=1.10 and the intercept of b=0.00 results in differences between the fitted line and the one-to-one line that are so small we chose not to apply any transformation to our $\rm S_{HK}$ measurements.  This close correspondence with the Mt. Wilson scale also indicates that the equation given in \cite{isaacson2010ApJ...725..875I}, with a V-flux coefficient adjusted to the values for each spectrum, yields values of $\rm S_{HK}$ that are close to the Mt. Wilson system.
We note that the star highlighted in parentheses in Figure \ref{fig:SHK_comp} (EPIC 212069861) is a M0-dwarf, one of the most active stars in our sample.
The S$\rm _{HK}$ measurement for this star in this study was derived from a single available Keck HIRES spectrum observed on November 11, 2016. Given the very large difference in the S$\rm _{HK}$ measurements for this star, it is likely that its activity level has varied significantly between the measurements here (S$\rm _{HK}=2.37$) and in \cite{chahal2023MNRAS.525.4026C} (S$\rm _{HK}$=1.43 $\pm$ 0.08). 
This star has not been included in the best fit slope shown in Figure \ref{fig:SHK_comp}.

\subsection{Conversion from S-index to R$\rm ^\prime _{HK}$}
The equation for the S-index includes terms with flux contributions from both the chromosphere plus the basal photosphere, which are referred to as $R^\prime_{\rm HK}$ and $R_{\rm phot}$, respectively. In order to isolate and measure chromospheric activity in stars with different effective temperatures, it is necessary to subtract the photospheric contribution by considering the dependence of the R and V fluxes on the effective temperature.  We adopted the method described in \citet{noyes1984ApJ...279..763N}, taking

\begin{equation}\label{eq:RpHK}
   R^\prime _{\rm HK} = R_{\rm HK} - R_{\rm phot} = 1.34 \times 10^{-4} C_{\rm cf} {\rm S_{HK}}  -  R_{\rm phot},
\end{equation}
where the photospheric contribution, $R_{\rm phot}$, and the temperature dependent bolometric correction factor, $C_{\rm cf}$, are modeled as polynomial functions of $(B-V)$:
\begin{equation}
   \log R_{phot} = -4.898 + 1.918(B-V)^2 - 2.893(B-V)^3.
\end{equation}
Since our sample includes FGK dwarfs and some evolved stars, we applied the bolometric corrections proposed by \cite{rutten1984AA...130..353R}, defined for 0.3 $\leq (B-V) \leq$ 1.7. 
 
For main sequence stars, taken as those stars falling below the black dashed line in Figure \ref{fig:kiel_sample}, we used:
\begin{equation}
    \log C_{\rm cf} = 0.25(B-V)^3 - 1.33 (B-V)^2 + 0.43 (B-V) + 0.24,
\end{equation}
while for evolved stars we used:
\begin{equation}
    \log C_{\rm cf} = -0.066(B-V)^3 - 0.25 (B-V)^2 - 0.49 (B-V) + 0.45.
\end{equation}
Finally, $R^\prime_{\rm HK}$ in Equation \ref{eq:RpHK} depends only on $(B-V)$, and the applicability of this relations is no longer limited only to late-F to early-K dwarfs. The mean $\log R^\prime_{\rm HK}$ obtained for our sample stars are presented in Table \ref{tab:parameters}. 

\section{Discussion} \label{sec:discus}

\subsection{Stellar Activity of the Sample K2 Targets}
\begin{figure*}
\gridline{\fig{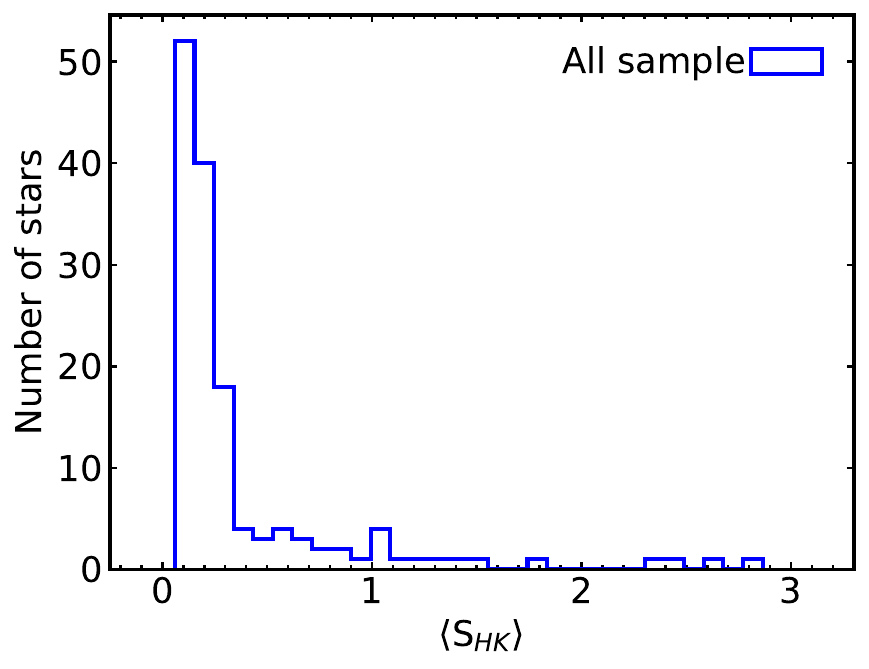}{0.48\textwidth}{}
            \hspace{-7mm}
          \fig{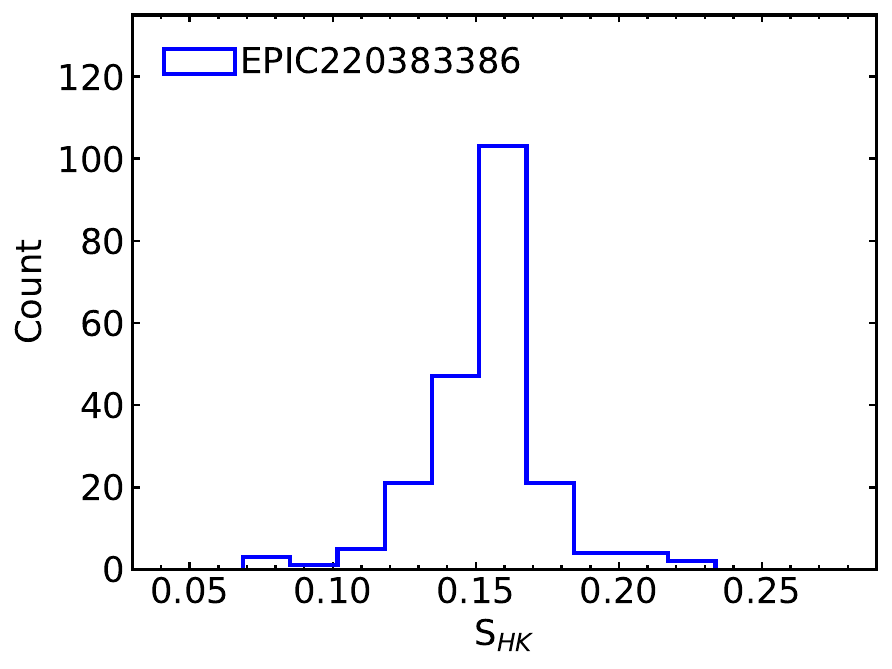}{0.492\textwidth}{}
          }
          \vspace{-7mm}
          \caption{Left panel: the distribution of the mean S$\rm _{HK}$  ($\rm \langle S_{HK} \rangle$) obtained for 111 K2 stars in our sample. Right panel: the S$\rm _{HK}$ index distribution for the star EPIC 220383386; this is the star in our sample with the largest number of HIRES spectra available: 211 spectra between July 10, 2016, and January 19, 2022.}    \label{fig:epic220383386}
\end{figure*}

In Section \ref{sec:S_ind} we discussed the derivation of the $\rm S_{HK}$ index, and in the left panel of Figure \ref{fig:epic220383386} we show the distribution of the derived indices for our sample of K2 stars. Most of the K2 targets in this study have $\rm S_{HK}$ $<$ 0.2, with the median of the distribution of $\rm S_{HK}= 0.178$ (16th percentile 0.134; 84th percentile = 0.577). There are, however, a few stars in our sample that have $\rm S_{HK}$ $\sim$ 2 -- 3, signifying high levels of activity.  The range of $\rm S_{HK}$-index values measured in this sample are similar to those presented in the large compilation of Mt. Wilson measurements from \cite{duncan1991ApJS...76..383D}.
In the right panel of Figure \ref{fig:epic220383386} we show the $\rm S_{HK}$ index distribution obtained for the star in our sample having the largest number of available HIRES spectra, EPIC 220725186. The $\rm S_{HK}$-index determinations for this star were based on 211 HIRES spectra collected over a period of six years (from July 10 2016 to January 19, 2022). The distribution of the S$\rm _{HK}$ indices for this star is peaked, having a mean S$\rm _{HK}$-index of $0.154 \pm 0.021$ (median of $0.158 \pm 0.009$), which indicates small stellar variability and its mean $\rm S_{HK}$-index is similar to the peak of the $\rm S_{HK}$-index distribution of our K2 sample.

It is useful for our discussion to note the S-index of the Sun within the distribution of $\rm S_{HK}$-index values shown in the left panel of Figure \ref{fig:epic220383386}. \cite{egeland2017ApJ...835...25E} used solar observations from both Mt. Wilson and Sacramento Peak to place the Sun on the Mt. Wilson $\rm S_{HK}$-index scale. Over solar cycles 15-24, \cite{egeland2017ApJ...835...25E} find a mean S-index of $\rm \langle S\rangle =0.1694\pm0.0020$, and at solar minima (the quiet Sun) to be $\rm \langle S_{min}\rangle =0.1621\pm0.0008$.  The variations in the S-index over the solar cycle were found to have a mean full amplitude of $\rm \langle\Delta S_{cyc}\rangle$=0.0145$\pm$0.0012. \cite{lorenzo2018AA...619A..73L} used ESO/HARPS spectra to measure the solar S-index from reflected light of solar system objects, in the same Mount-Wilson scale as that used for their sample of solar twins observed through the same spectrograph, and found an average solar activity of $\rm \langle S_{MW}\rangle =0.17$. Using a correlation between activity and sunspots, they estimated an average S-index for the solar cycles 15 - 24 of $\rm \langle S_{MW}\rangle=0.1696\pm0.0025$, in perfect agreement with the same cycles analyzed by \citep{egeland2017ApJ...835...25E}.
Since most of the stars in our sample have only one, or a few observations, our $\rm S_{HK}$ measurements present a ``snapshot'' of chromospheric activity over the sample of stars which were caught in random phases over their respective activity cycles and, thus, we are not in a position to discuss individual stellar magnetic cycles nor rotational period coverage.  Nonetheless, our results yield a distribution of $\rm S_{HK}$ values sampled at random over a population of FGK stars and we now discuss some characteristics of this sample.

\subsubsection{S$_{\rm HK}$ vs. $(B-V)$}
\begin{figure*}
    \centering
    \includegraphics[scale=0.5]{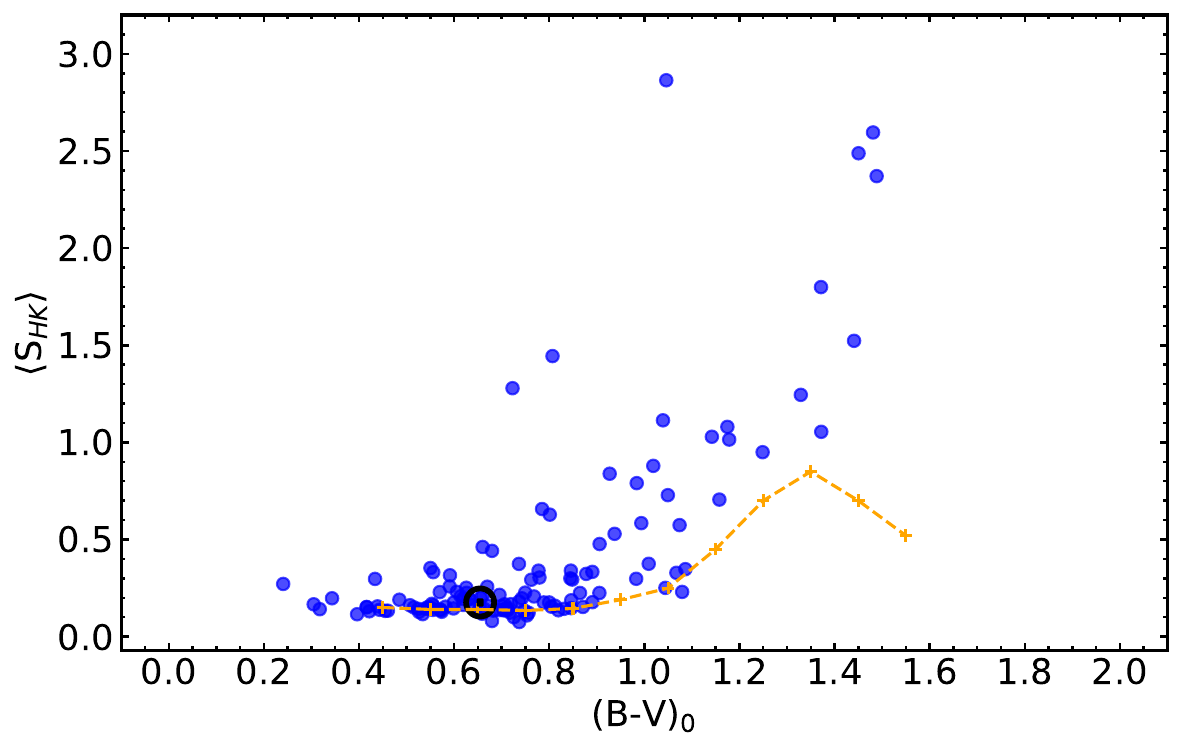}
    \caption{Mean S$\rm _{HK}$ indices as a function of the $(B-V)_{0}$ color for the main sequence stars in our K2 sample. The solar symbol represents the solar value at  $\rm\langle S_{HK}\rangle=0.176$ \citep{mamajek2008ApJ...687.1264M}, lying close to the dashed orange line taken from \cite{isaacson2010ApJ...725..875I}.}
    \label{fig:SHK_BV}
\end{figure*}

Figure \ref{fig:SHK_BV} shows the $\rm S_{HK}$ for the K2 targets measured here as a function of their $(B-V)_{0}$ color. 
Both \cite{isaacson2010ApJ...725..875I} and \cite{boro_saikia2018AA...616A.108B} observed an increase in the S-index for $(B-V)$ colors ranging from $\sim$ 1.0 to 1.4 in a sample of $\sim2600$ and $\sim4400$ main-sequence stars, respectively.  The lower envelope increase of $\rm S_{HK}$ with increasing $(B-V)_{0}$ is due primarily to the decreasing flux in the V and R continuum fluxes in cooler stars relative to the fluxes in the Ca II H and K lines \citep[e.g.,][]{isaacson2010ApJ...725..875I}.
In this study of 144 K2 stars within the same range in $(B-V)_{0}$, our results for S-index vs $(B-V)_{0}$ shown in the figure also exhibit a similar pattern for $(B-V) < 1.4$, as the S-index begins to increase roughly at $(B-V)_{0}\sim 0.8$. 
The general agreement is made clearer in comparison with the orange dashed line in the figure, which represents the polynomial fit, taken from \cite{isaacson2010ApJ...725..875I}, and traces the lower envelope of their distribution.  

In our sample the lower activity limit remains approximately constant for $(B-V)_{0}\lesssim 0.9$, while for redder colors in the range of $0.9 \lesssim (B-V)_{0} \lesssim 1.4$, the lower limit of the S-index exhibits a smooth increase, from $\rm S_{HK}$ $\sim$0.2 at $(B-V)_{0}=0.9$ up to $\rm S_{HK}$ $\sim$0.8 at $(B-V)_{0}=1.4$, and begins to decrease for $(B-V)_{0} \gtrsim 1.4$. Notably, throughout this entire range, the highest levels of activity, or the fraction of active stars, increase sharply towards redder values of $(B-V)_{0}$ (later spectral types). 
For colors of $(B-V)_{0}<0.9$ there are only  8 \rm stars out of 30 stars with $\rm S_{HK} > 0.4$ (roughly twice the value of the lower envelope), while in the range of $(B-V)_{0}\sim 0.9$ to 1.2 there are 14 stars out of 22 that exhibit $\rm S_{HK}$ values that are more than twice the lower envelope.

\subsubsection{$R^{\prime}_{\rm HK}$ and Stellar Rotation} 
\begin{figure}[!h]
    \centering
    \includegraphics[scale=0.57]{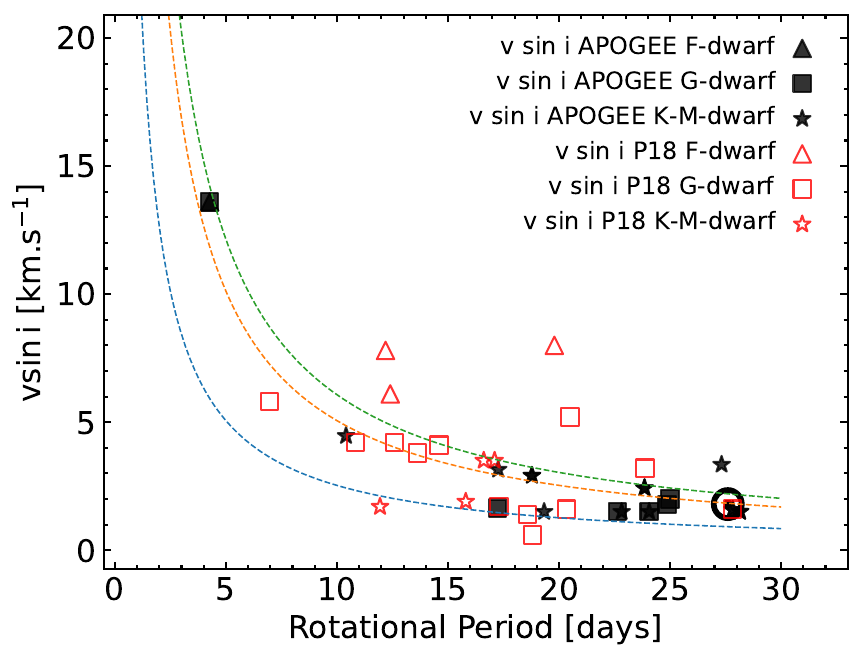}
    \caption{Projected rotation velocities ($v \sin i$) and measured rotation periods for those stars in our sample having the two measurements available in the literature. The different symbols segregate the stars according to their spectral types. The three dashed lines represent the equatorial rotational velocity as a function of rotational period between 0.1 to 30 days, for the following stellar radii: 0.5 R$_\odot$ (blue), 1.0 R$_\odot$ (orange), and 1.2 R$_\odot$ (green).}
    \label{fig:vsin_Prot}
\end{figure}

As a star evolves it loses angular momentum, resulting in lower stellar rotational velocities, which is accompanied by a decrease in stellar activity \citep{skumanich1972ApJ...171..565S}. 
In order to analyze the behavior of the activity index $R^\prime_{\rm HK}$ in relation to $v \sin i$ and stellar rotational period, we collected $v \sin i$ measurements for our sample stars from the DR17 APOGEE survey \citep{majewski2017} and \cite{petigura2018}, these values are presented in the last column of the Table \ref{tab:parameters}.  
The $v \sin i$ sample is a mixture of Keck/HIRES spectra, with R $\sim$ 60,000, and APOGEE spectra, with R $\sim$ 22,400, providing theoretical lower limits to $v \sin i$ detections of roughly $\sim 2-3$ km-s$^{-1}$. The rotational periods for the sample were taken from \cite{reinhold2020AA...635A..43R}, \cite{deleon2021MNRAS.508..195D}, and \cite{rampalli2021ApJ...921..167R}, however rotational periods were available for only 44 sample stars. 
In Figure \ref{fig:vsin_Prot} we plot $v \sin i$ versus rotational period for those stars that have both measurements available, using different symbols for stars as a function of their spectral types and also identify the $v \sin i$ measurement source as either from APOGEE or \cite{petigura2018}.  There is an expected increase in $v \sin i$ as the rotational period decreases and the star having the highest $v \sin i$ in this sample ($v \sin i$ = 13.6 km/s) has a measured rotational period of less than 5 days. These independent measurements are self-consistent: a G-dwarf having an approximate radius of 1 R$_{\odot}$ and a rotational period of $\sim$4.5 days would have an equatorial rotational velocity of $\sim$12 km-s$^{-1}$. The behavior of the equatorial rotational velocity as a function of rotational period for 0.5 R$_\odot$, 1.0 R$_\odot$, and 1.2 R$_\odot$ is shown in Figure \ref{fig:vsin_Prot} as the blue, orange and green dashed line, respectively. We can identify a floor in the $v \sin i$ at around $2-3$ km-s$^{-1}$ which is likely due to the loss in sensitivity given the intrinsic broadening in stellar spectra due to microturbulence and macroturbulence.

\begin{figure*}
    \centering
    \includegraphics[scale=0.475]{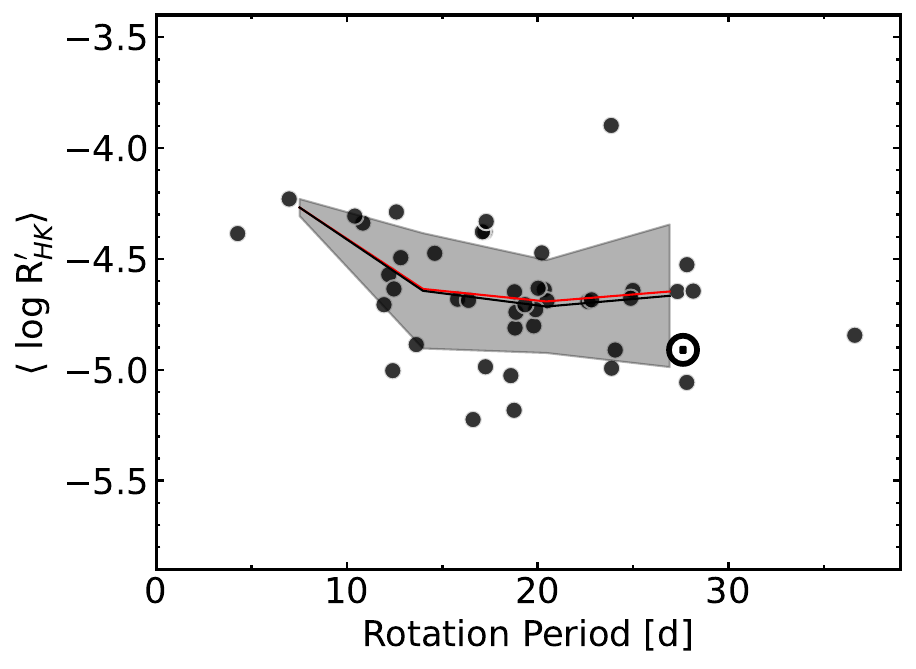}
    \includegraphics[scale=0.48]{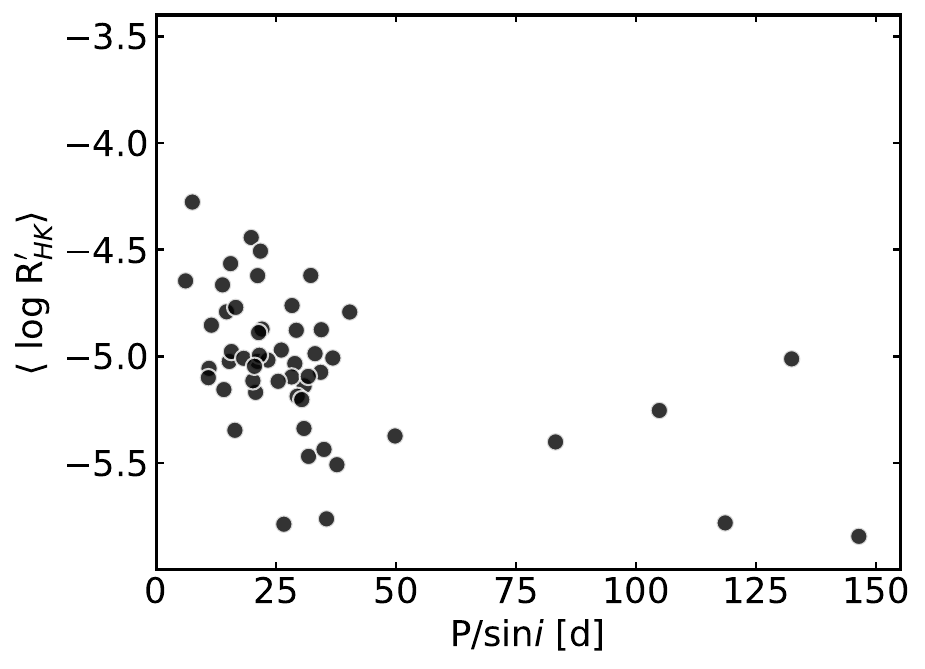}
    
    \includegraphics[scale=0.475]{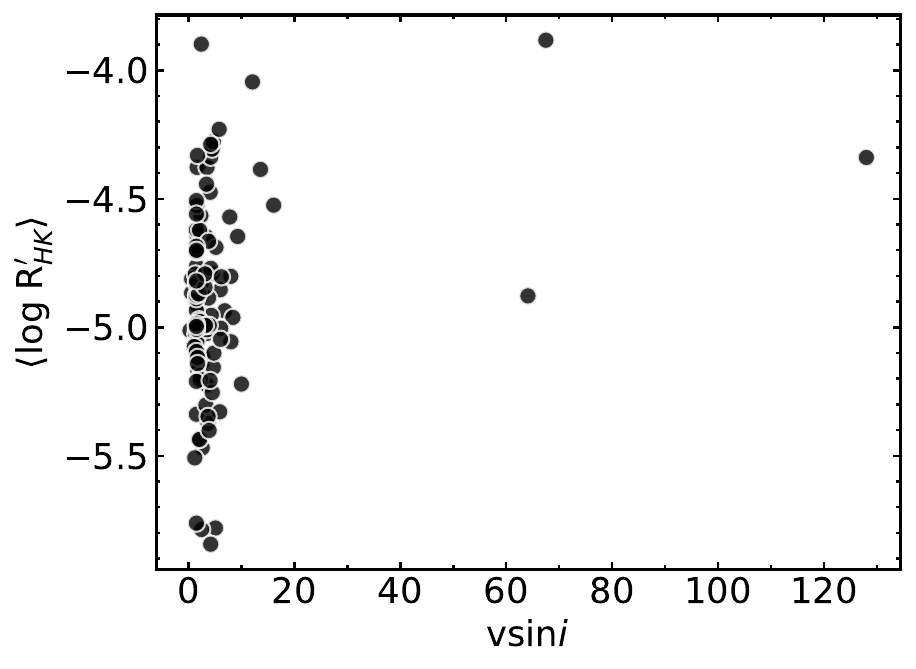}
    \includegraphics[scale=0.475]{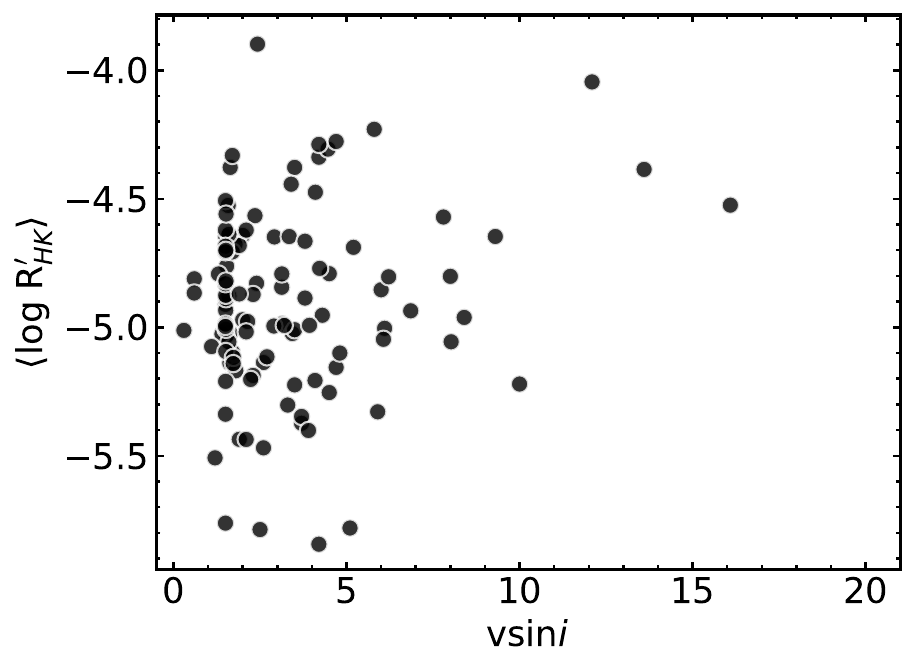}
    \caption{Top right panel:   Top right panel shows mean $\log R^\prime_{\rm HK}$ index as a function of rotational period. The red line represents the running median and the black line the running mean, the gray area is the standard deviation. Bottom panels: mean $\log R^\prime_{\rm HK}$ index as a function of rotational projected velocity ($v \sin i$).}
    \label{fig:SHK_vsini}
\end{figure*}

The top left panel of Figure \ref{fig:SHK_vsini} shows $\langle \log R^\prime_{\rm HK} \rangle$ versus the rotational period for the 44 stars in our sample with measured rotational periods available in the literature, as discussed earlier in this section.  
The distribution of stars with measured rotational periods for our sample is skewed towards higher levels of activity, with only two stars having $\langle$log R$^\prime _{\rm HK}\rangle <-5.10$. This is likely due to more active stars having stellar surface features, such as spots, that can be detected via K2 photometry.
Nevertheless, there is a trend in this sub-sample, as illustrated by the red running median and black running mean lines, in which the distribution of $\langle \log R^\prime _{\rm HK}\rangle$ values remains nearly flat at longer rotational periods ($\rm P\sim 14-30$ days) and then increases for lower rotational periods (higher rotational velocities). 
Our results agree well with measurements of $\langle \log R^\prime_{\rm HK} \rangle$ versus rotational period for a sample of planet-hosting stars in the Kepler prime field by  \cite{morris2017ApJ...848...58M}. 
As a reference, a period of 12 days for a solar-radius star would have an equatorial rotational velocity of $\sim 4-5$ km-s$^{-1}$.

To further investigate trends between rotation and activity, we derived P/$\sin i$ from the $v \sin i$ values and these are shown in the top right panel of Figure \ref{fig:SHK_vsini}, noting that those stars with measured rotational periods have been excluded from this figure.  The stars with very large values of P/$\sin i > 80$ days are evolved stars (see Figure \ref{fig:kiel_sample}), with the main sequence stars having P/$\sin i \lesssim$ 40 days, keeping in mind that these periods are upper limits, with typical expected inclination effects of $\sim0-40$\% for half of the sample. Unlike the upper left panel of this figure, with measured photometric periods, this sample exhibits the full range of activity levels and with large scatter in $\log R^\prime_{\rm HK}$ at a given value of P/$\sin i$. Within this scatter, however, both the lower and upper envelopes define increasing $\log R^\prime_{\rm HK}$ with decreasing P/$\sin i$ and the points from the upper left panel are contained within the high activity region of the envelope.  
Although there is not a simple relation in the upper panels of Figure \ref{fig:SHK_vsini}, the general trend is that stars show more activity for low values of P/$\sin i$ with a steep rise in $\langle \log R^\prime_{\rm HK}\rangle$ beginning at approximately $P \sin i$ $\sim$ 30--40 days. The stars with P/$\sin i$ higher than this threshold all show very low levels of activity at $\log R^\prime_{\rm HK} \sim -5.5$.
The lack of a simple relation between $\log R^\prime_{\rm HK}$ as a function P$_{rot}$ is also found by \cite{bohmvitense2007ApJ...657..486B} in a study of 25 G and K dwarfs which display large scatter in $\log R^\prime_{\rm HK}$ at a given rotational period, with the suggestion of two sequences that \cite{bohmvitense2007ApJ...657..486B} labels as active (A) and inactive (I), although some stars are found in-between the two sequences.  Similar results are found by \cite{metcalfe2017SoPh..292..126M} in a sample of $\sim$50 FGK dwarfs. The results from both \cite{bohmvitense2007ApJ...657..486B} and \cite{metcalfe2017SoPh..292..126M} overlap closely our results as presented in the top panels of Figure \ref{fig:SHK_vsini}, over the rotational period range of $\sim$5--45 days. To better illustrate this comparison, the approximate envelope in $\log R^\prime_{\rm HK}$ versus P$_{rot}$ found by \cite{metcalfe2017SoPh..292..126M} in their Figure 2 is outlined by the solid lines in the top right panel of Figure \ref{fig:SHK_vsini}, while the dashed lines follow the respective Inactive and Active sequences from \cite{bohmvitense2007ApJ...657..486B} in her Figure 8. These two earlier studies, along with the results presented here, point to a complex interplay between stellar rotational period and chromospheric activity that is likely a function of both stellar mass and age.

The bottom panels of Figure \ref{fig:SHK_vsini} present the literature $v \sin i$ measurements along with their corresponding $\log R^\prime_{\rm HK}$ values derived in this study, with the left panel showing all stars, while the right panel has an expanded scale showing $v \sin i$ values up to 20 km-s$^{-1}$. Given that the $v \sin i$ values represent lower limits to the true rotational velocities, stars with low values of $v \sin i$ $< 3$ km-s$^{-1}$ (the approximate floor in detection limit from Figure \ref{fig:vsin_Prot}) are composed of a mixture of slowly rotating stars and rapidly rotating stars that are viewed at small inclination angles. Stars with higher values of $v \sin i$ 
can be considered as bone fide rapid rotators. 
As shown in the right bottom panel of Figure \ref{fig:SHK_vsini}, the stars with $v \sin i$ $<$ 3 km-s$^{-1}$, which are a mixture of slow and rapid rotators, exhibit a large range in $\log R^\prime_{\rm HK}$ and contain 20 stars having high activity levels of $\log R^\prime_{\rm HK} \ge -4.75$, or 14.3\%  being in the high activity group. As discussed above, it is likely that a fraction of the stars in this sub-sample are fast rotators viewed at small inclination angles. There remain five stars with $v \sin i$ $>$10 km-s$^{-1}$ and this sub-sample rapid rotators contains a larger fraction of 83\% (5 out of 6) active stars, with $\log R^\prime_{\rm HK}\ge -4.75$, supporting the general result that rapidly rotating stars tend to have higher levels of chromospheric activity.

\subsubsection{The Vaughan-Preston gap}
\cite{vaughan1980PASP...92..385V} used the Ca II H and K lines to map chromospheric activity in 486 field stars in the solar neighborhood as part of the Mount Wilson project \citep{wilson1968ApJ...153..221W,duncan1991ApJS...76..383D,baluinas1995ApJ...438..269B}. They found that in this sample there was a lack of F- and G-type stars with intermediate activity levels, which has been labeled the ``Vaughan-Preston gap'', with this gap spanning a range in $(B-V)_{0}$ roughly between 0.4 and 1.1 and values of S$_{\rm HK}$ between 0.2 to 0.7. 
A lack of stars in this regime was later confirmed by \cite{henry1996AJ....111..439H} from an analysis of 800 stars observed with the Coude' spectrograph on the 1.5 m Telescope at CTIO and more recently by \cite{gomes2021AA...646A..77G} from analysis of HARPS spectra of 1674 FGK stars.  \cite{boro_saikia2018AA...616A.108B} also investigated stellar activity in a large sample of cool stars using results collected from various surveys in the literature and concluded that the Vaughan-Preston gap may not be a significant feature. 

\begin{figure}
    \includegraphics[scale=0.48]{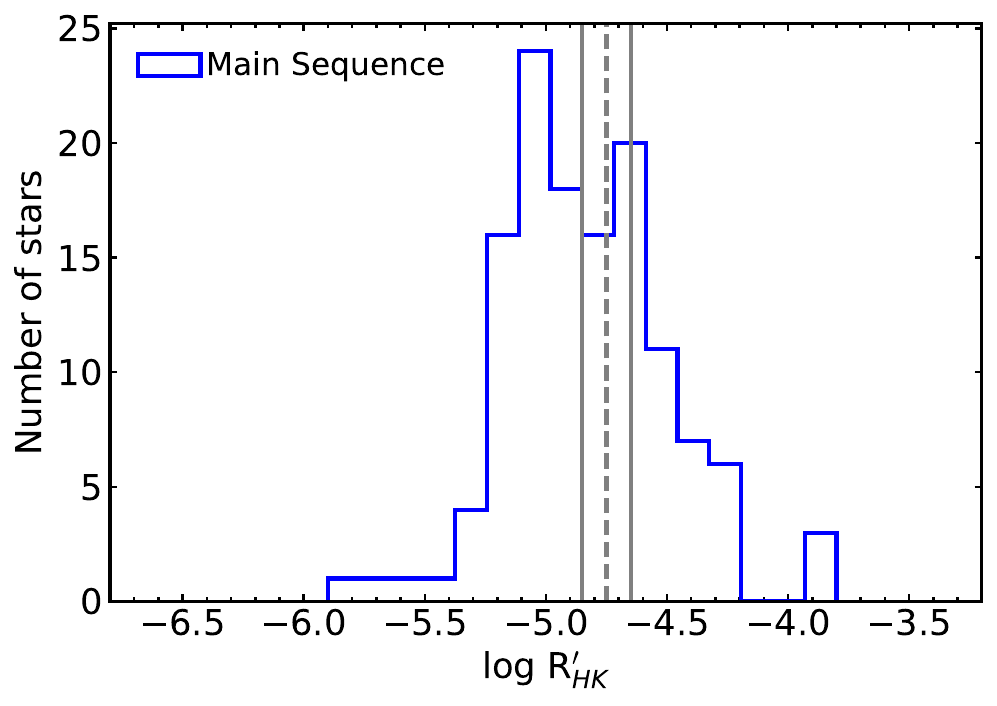}
    
    \includegraphics[scale=0.48]{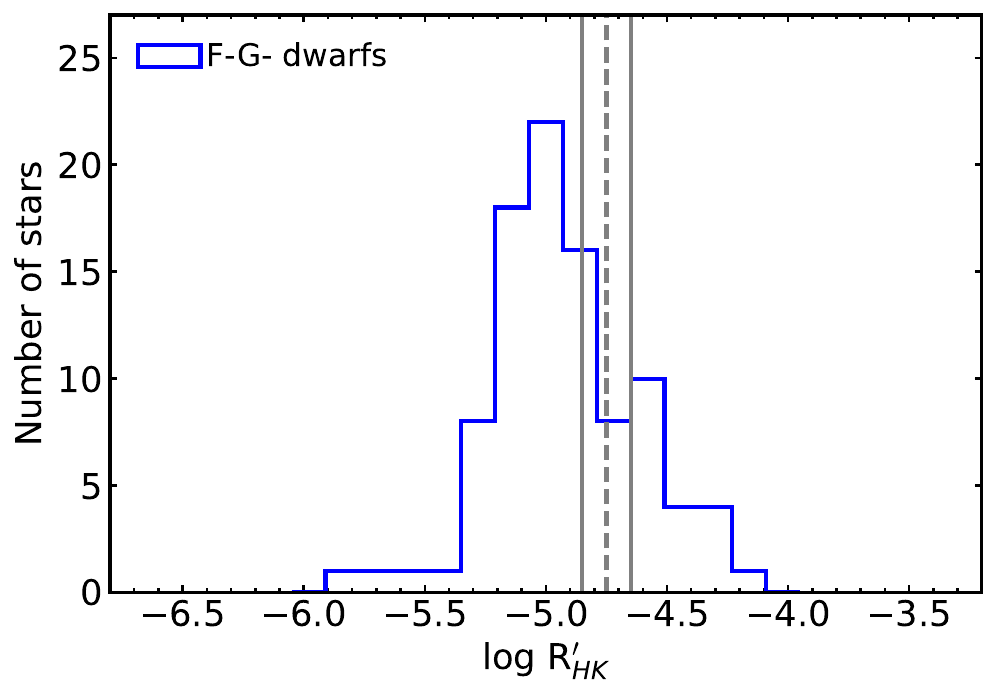}
    
    \includegraphics[scale=0.49]{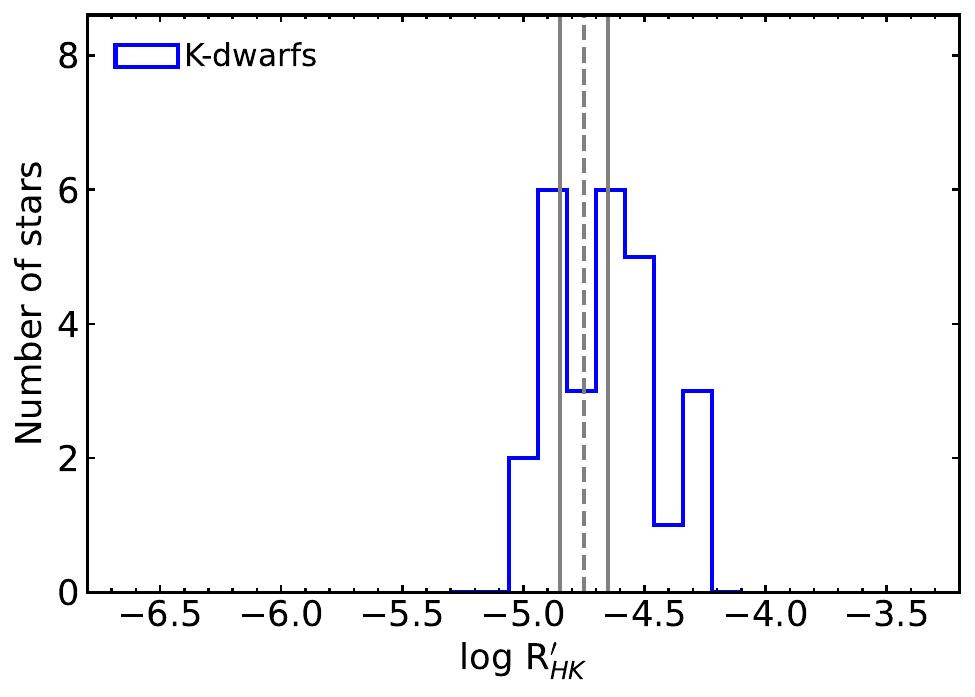}
    \caption{Top panel: the distribution of $\log R^\prime_{\rm HK}$ for the main sequence stars in our K2 sample.  Middle panel: the distribution of $\log R^\prime_{\rm HK}$ for the F-G- type dwarfs in our sample. Bottom panel: the distribution of $\log R^\prime_{\rm HK}$ for the K-type dwarfs in our sample.The region of the Vaughn-Preston gap is marked by the gray solid lines in all panels.} 
    \label{fig:logRHK_dist_gap_BV}
\end{figure}

The top panel of Figure \ref{fig:logRHK_dist_gap_BV} shows the distribution of the mean values of $\log R^\prime_{\rm HK}$ for our sample of main sequence stars and reveals a gap, or valley, with a depth of $\sim$35\% of the main peak, at the approximate location of the Vaughan-Preston gap (shown as the vertical lines), with boundaries as defined in \citet{boro_saikia2018AA...616A.108B}. This boundary delineates a bimodal distribution around the gap.
In the middle panel of the figure we show the $\log R^\prime_{\rm HK}$ distribution only for the sub-sample of F and G-type dwarfs, using for this selection a threshold of T$_{\rm eff}$ $>$ 5000 K. Although this distribution is not clearly bimodal, it is noted that the peak at the side of the gap corresponding to active stars is smaller when compared to the distribution shown in the top panel of Figure \ref{fig:logRHK_dist_gap_BV}. 
\cite{gomes2021AA...646A..77G} noted bimodal distributions in the values of $\log R^\prime_{\rm HK}$ for their sample of F and G dwarfs with means close to --4.95 and --4.50 for the inactive and active stars, respectively, which are close to ours. 
That study, with a much larger number of stars, found overall a similar distribution to ours, with a similarly smaller peak on the active side of the gap when compared to the non-active side. Since \cite{gomes2021AA...646A..77G} had a much larger sample than the one we present here, they carried out two, three, four, and five component Gaussian fits to their results, finding that three and four component models yield somewhat better fits to their results; our sample is not large enough to meaningfully conduct such tests. 
\cite{gomes2021AA...646A..77G} also found that the K-dwarfs in their sample had a different distribution of $\log R^\prime_{\rm HK}$ than the F and G dwarfs and exhibited three distinct peaks, with no Vaughan-Preston gap. Although our sample is dominated by F and G dwarfs, having only 26 K dwarfs, in the bottom panel of Figure \ref{fig:logRHK_dist_gap_BV} we present the $\log R^\prime_{\rm HK}$ distribution of the K dwarfs. Given the small number of stars in this sub-sample it is hard to conclude that we also see three peaks in the distribution of K dwarfs, although the region of the Vaughan-Preston gap shows a reduced number of stars 
with the obvious caveat that we have very few K-dwarfs in our sample. One feature of the distribution of these stars is that they all have $\log R^\prime_{\rm HK} > -5$, none of them belonging to the ``very inactive stars'' category.

\subsubsection{The Impact of Including Activity Sensitive Lines in the Analysis}\label{sec:impact}

In Section \ref{sec:st_par} we described our derivation of stellar parameters that is based upon Fe I and Fe II lines and which relied only on lines deemed to be insensitive to stellar magnetic activity, as described in \cite{yana2019MNRAS.490L..86Y}. In this section we investigate possible differences between the stellar parameters derived without using activity-sensitive Fe lines (see Section \ref{sec:st_par}) when compared to those derived with the inclusion of a significant number of activity-sensitive Fe lines and check whether any differences in $\Delta T_{\rm eff}$, $\Delta$[Fe/H], or $\Delta \xi$ correlate with $\log R^\prime_{\rm HK}$.
A number of recent studies using high-resolution spectra ($\rm R \sim 115,000$), such as those by \cite{flores2016A&A...589A.135F}, \cite{lorenzo2018AA...619A..73L}, \cite{yana2019MNRAS.490L..86Y}, and \cite{spina2020ApJ...895...52S}, have shown that stellar magnetic activity can affect determinations of $T_{\rm eff}$, $\xi$, and [Fe/H] that are based upon an analysis using Fe I and Fe II lines.  Figures 3 and 4 in \cite{spina2020ApJ...895...52S} show that there can be an enhancement in the strengths of certain magnetically sensitive Fe I lines in stars with $\log R^\prime_{\rm HK} > -5.0$. As reported by \cite{spina2020ApJ...895...52S}, the presence of stellar activity may lead to differences in the derived effective temperature, metallicity, and microturbulent velocity, but with no significant variations in the surface gravity determinations.  In particular, they found a decrease in the effective temperatures for stars with $\log R^\prime_{\rm HK} > -4.6$, while a decrease in [Fe/H] values begins at $\log R^\prime_{\rm HK} \sim -5.0$, along with an increase in the microturbulent velocity. In addition, \cite{yana2019MNRAS.490L..86Y} analyzed several observations of the solar twin star HIP 36515 and showed that the difference in the derived parameters computed with and without activity-sensitive lines varies depending on the phase of the stellar activity cycle. Here we will investigate the impact in the stellar parameter determinations from inclusion of an additional 49 Fe I lines and 2 Fe II lines that were found to be sensitive to stellar activity.

\begin{figure*}
    \centering
    \epsscale{1.25}
    \plotone{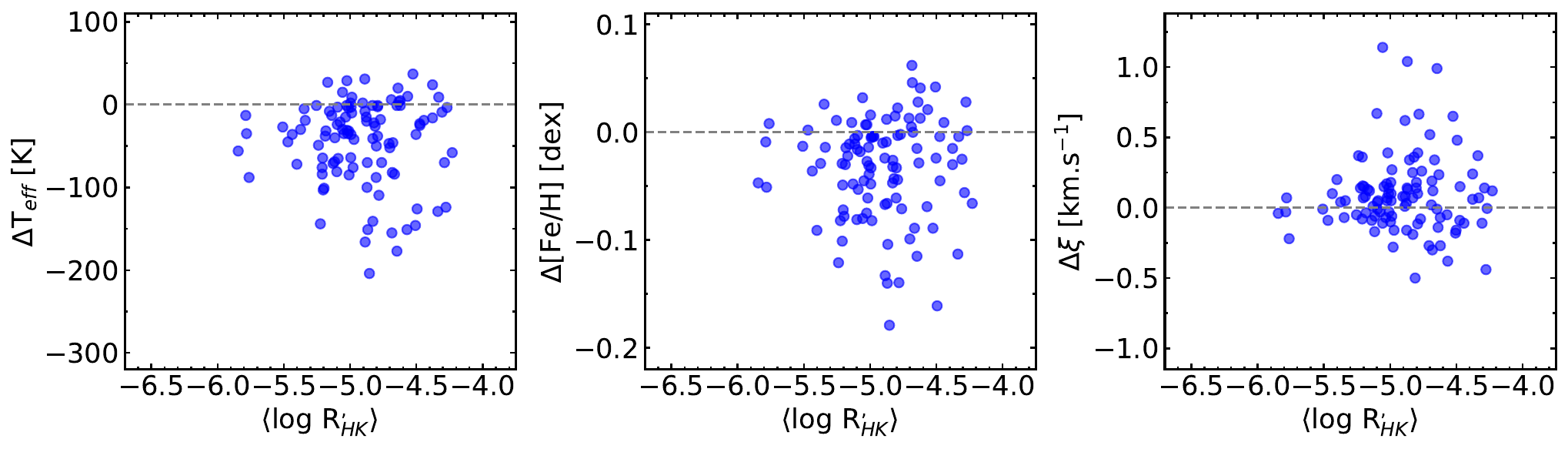}
    \caption{Differences ($\Delta$) between effective temperatures, metallicities and microturbulent velocities determined without including Fe lines which are sensitive to activity (This study) and those determined including a significant number of activity sensitive Fe lines, as a function the mean activity index $\log R^{\prime}_{\rm HK}$ measured for the stars.}
    \label{fig:Delta_SP}
\end{figure*}

As shown in Figure \ref{fig:Delta_SP} there are systematic differences between the results in the sense that the effective temperatures and metallicities derived including activity sensitive Fe lines are overall larger than those derived without sensitive lines. The mean differences (``non sensitive $-$ sensitive'')  and standard deviations are: $-46 \pm 51$ K, $-0.034 \pm 0.047$, and $0.099 \pm 0.277$ km.s$^{-1}$ for $T_{\rm eff}$, metallicity, and microturbulent velocity, respectively. However, these are small differences which are of the order of the uncertainties in the derived parameters (see Table \ref{tab:error_budget}), which leads us to conclude that, based on the sample of Fe I and Fe II lines used here, there is no evidence (beyond the uncertainties) of a clear correlation between $\Delta T_{\rm eff}$, $\Delta$[Fe/H] and $\Delta \xi$ with $\log R^{\prime}_{\rm HK}$. 

Furthermore, if we segregate this sample into 3 regimes in terms of stellar activity \citep[from][]{henry1996AJ....111..439H} we find that the mean $T_{\rm eff}$, [Fe/H] and $\xi$ differences for the sub-samples of very inactive, inactive, and active stars are respectively $-$49 K, $-$0.035 dex, 0.063 km.s$^{-1}$; $-$40 K, $-$0.038 dex, 0.128 km.s$^{-1}$; and $-$47 K, $-$0.024 dex, 0.068 km.s$^{-1}$, again not showing a clear signature with $\log R^{\prime}_{\rm HK}$. 
The only signature to note is that the spread in $\Delta T_{\rm eff}$ and $\Delta$[Fe/H] increases for $\log R^{\prime}_{\rm HK} >\sim -5$. The 8 stars with the largest differences in $T_{\rm eff}$ ($>\sim$110 K) all have $\log R^{\prime}_{\rm HK}$ larger than $\sim-$5.1. This is also the case for the metallicity results shown in the middle panel of Figure \ref{fig:Delta_SP}: those stars with the largest variations in metallicity have $\log R^{\prime}_{\rm HK}$ larger than $\sim-$5. 
In summary, very inactive stars display smaller spreads in $\Delta T_{\rm eff}$ and $\Delta$[Fe/H] when compared to the more active stars.

\section{Summary and Conclusions} \label{sec:conclusion}
The results presented here were obtained through a spectroscopic analysis of a uniform set of high-resolution Keck I/HIRES spectra using a list of Fe I and Fe II lines, which yielded the fundamental stellar parameters of $T_{\rm eff}$, $\log g$, $\xi$, and [Fe/H], as well as stellar radii. The analysis employed the spectroscopic analysis code $q^2$ \citep{Ramirez2014}, which derives stellar parameters semi-automatically.  
The sample includes 109 K2 stars, of which 75 have confirmed planets; the derived values of stellar radii were used to determine planetary radii. The Fe I and Fe II lines used in the analysis were from \cite{yana2019MNRAS.490L..86Y} and consists of some lines that are sensitive to magnetic fields, as well as some lines that are insensitive.  Main results are summarized below.

$\bullet$ Our stellar radii were combined with transit depths to calculate planetary radii for 93 confirmed planets orbiting 69 K2 stars.  Most of transit depths were obtained from \cite{kruse2019}, with additional transit depths from \cite{barros2016AA...594A.100B}, \cite{pope2016MNRAS.461.3399P}, and \cite{christiansen2017AJ....154..122C}. The internal precision of the planetary radii achieved here was 2.3\% and the radius gap is detected in this sample of K2 planets, located at R$_{pl}$ $\sim$ 1.9 R$_\oplus$.

$\bullet$ Stellar activity using the $\rm S_{HK}$ index, which is defined by the fluxes of the Ca II H and K lines, was measured for 144 stars using spectra obtained from ExoFop and the KOA. 
The values of S$\rm _{HK}$ were derived using the equation in \cite{isaacson2010ApJ...725..875I}, with a V-flux coefficient fit to values obtained for each spectrum, and were found to fall closely on the Mt. Wilson scale. Our derived values of $\rm S_{HK}$ as a function of $(B-V)_0$ follow the same trends as shown in \cite{isaacson2010ApJ...725..875I} and \cite{boro_saikia2018AA...616A.108B}.
Photospheric contributions to $\rm S_{HK}$ were subtracted to obtain values of $\log R^{\prime}_{\rm HK}$ by using bolometric corrections from \cite{rutten1984AA...130..353R} for main sequence and evolved stars. 

$\bullet$ Our results for $\langle \log R^\prime _{\rm HK}\rangle$ as a function of rotational period for the K2 sample indicate that although activity, in general, decreases with increasing rotational period, there is a large scatter in the activity level for a given rotational period of $\sim$0.5 dex in $\log R^\prime_{\rm HK}$; such a result has been noted previously, e.g., \cite{bohmvitense2007ApJ...657..486B} and \cite{metcalfe2017SoPh..292..126M}, and points to a complex behavior in stellar activity that is related possibly to Rossby number \citep{metcalfe2017SoPh..292..126M}. 

$\bullet$ In a previous study of chromospheric activity in a sample of solar-type stars, \cite{vaughan1980PASP...92..385V} found a lack of F- and G-type stars with a intermediate activity levels, now called the ``Vaughan-Preston gap''. This gap was confirmed by \cite{henry1996AJ....111..439H} and \cite{gomes2021AA...646A..77G} using a large sample of stars. Although we studied a much smaller sample of stars, our distribution of the mean values of $\log R^\prime_{\rm HK}$ also reveal the presence of the Vaughan-Preston gap.

$\bullet$ The possible impact of stellar activity on the derivation of the stellar parameters $T_{\rm eff}$, [Fe/H], and $\xi$ (surface gravity is not expected to be sensitive to stellar activity) was investigated by including magnetically sensitive Fe I lines in a separate spectroscopic analysis. We found systematic offsets between the two analyses with mean differences of ``non sensitive -- sensitive'' of $\Delta T_{\rm eff}=-46 \pm 51$ K, $\rm \Delta[Fe/H]=-0.03\pm0.05$, and $\rm \Delta\xi=+0.10\pm0.28$ km-s$^{-1}$. Although these systematic offsets are not significant and within the expected uncertainties, we note that the more active stars, with $\log R^\prime _{\rm HK} >-5.0$, exhibit larger scatter in their distributions of $\Delta T_{\rm eff}$ and $\Delta$[Fe/H] when compared to the quiet stars.



We thank the referee for detailed comments that helped improve the paper. V. L-T. acknowledges the financial support from Coordena\c{c}\~{a}o de Aperfei\c{c}oamento de Pessoal de Nível Superior (CAPES). This research has made use of the NASA Exoplanet Archive, which is operated by the California Institute of Technology, under contract with NASA under the Exoplanet Exploration Program.


\bibliography{manuscript}{}
\bibliographystyle{aasjournal}
\end{document}